\documentclass[prd,showpacs,floatfix]{revtex4}
\usepackage{epsfig}
\newcommand{\eq}[1]{Eq.~(\ref{#1})}
\newcommand{\fig}[1]{Fig.~{\ref{#1}}}
\newcommand{\be}{\begin{equation}}
\newcommand{\ee}{\end{equation}}
\newcommand{\bea}{\begin{eqnarray}}
\newcommand{\eea}{\end{eqnarray}}
\newcommand{\DS}{Dyson--Schwinger }
\newcommand{\w}{\omega}
\newcommand{\e}{\varepsilon}
\newcommand{\al}{\alpha}
\newcommand{\ba}{\beta}
\newcommand{\ga}{\gamma}
\newcommand{\G}{\Gamma}
\newcommand{\de}{\delta}

\newcommand{\si}{\sigma}

\newcommand{\ka}{\kappa}
\newcommand{\ha}{\frac{1}{2}}
\newcommand{\pd}{\partial}

\newcommand{\cd}{{\cal D}}
\newcommand{\cs}{{\cal S}}
\renewcommand{\div}{\vec{\nabla}}
\newcommand{\s}[2]{{#1}\!\cdot\!{#2}}
\newcommand{\ov}[1]{\overline{#1}}
\newcommand{\dk}[1]{\,\,\,\raisebox{-0.4ex}{\large $\bar{}$}\!\!d\,{#1}\,}

\newcommand{\ev}[1]{<\!\!{#1}\!\!>}
\newcommand{\kslash}{k\hspace{-2mm}\slash}

\begin{document}
%-----------------------------------------------------------------------------
\title{Quarks in Coulomb gauge perturbation theory}
\author{C.~Popovici}
\author{P.~Watson}
\author{H.~Reinhardt}
\affiliation
{Institut f\"ur Theoretische Physik, Universit\"at T\"ubingen, Auf der 
Morgenstelle 14, D-72076 T\"ubingen, Deutschland}
\begin{abstract}
Coulomb gauge quantum chromodynamics within the first order functional 
formalism is considered.  The quark contributions to the Dyson-Schwinger 
equations are derived and one-loop perturbative results for the two-point 
functions are presented.
\end{abstract}
\maketitle
%-----------------------------------------------------------------------------
\section{Introduction}

For more than thirty years, quantum chromodynamics [QCD] has been 
established as the theory of the strong interaction, yet fundamental 
questions still remain unanswered.  Foremost amongst these are the issues 
of confinement and the details of the hadron spectrum.  As was recognized 
early on, studies in Coulomb gauge (where the framework is naturally 
connected to physical degrees of freedom) should provide for a promising 
arena of endeavor.  However, what was also recognized at the time was that 
the inherent noncovariance of Coulomb gauge made technical progress 
difficult \cite{Abers:1973qs}.  This is one reason why the linear covariant 
gauges (such as Landau and Feynman gauges) have been preferred for both 
perturbative and nonperturbative calculations.  Clearly though, since the 
physical world is gauge invariant, it is certainly worthwhile to study the 
different gauges to gain more insight.  Perturbation theory is one area 
where results in different gauges can be unambiguously compared.  In 
addition, perturbative results form the basis of many nonperturbative 
studies since they provide a reliable way of dealing with ultraviolet 
divergences.

In the last few years, Coulomb gauge studies have been enjoying significant 
progress.  There exists an appealing picture for confinement: the 
Gribov--Zwanziger scenario 
\cite{Gribov:1977wm,Zwanziger:1995cv,Zwanziger:1998ez} whereby the temporal 
component of the gluon propagator provides for a long-range confining force 
whilst the transverse spatial components are suppressed in the infrared (and 
therefore do not appear as asymptotic states).  Recent lattice studies seem 
to confirm this \cite{Burgio:2008jr}.  A Hamiltonian-based approach \cite{Christ:1980ku,Szczepaniak:2001rg,Szczepaniak:2003ve,Feuchter:2004mk,
Reinhardt:2004mm} also exists and proves adept at describing various 
physical features of the system 
\cite{Reinhardt:2007wh,Reinhardt:2008ek,Campagnari:2008yg}.  More pertinent 
to this study, results in the Lagrange-based (Dyson--Schwinger) functional 
integral approach (which is especially suitable for perturbative 
calculations) to Yang--Mills theory have recently become available \cite{Watson:2006yq,Watson:2007mz,Watson:2007vc,Reinhardt:2008pr,Watson:2007fm}.

In this work, we consider the quark contributions in the first order 
functional approach to Coulomb gauge QCD, using results obtained in the pure 
Yang--Mills theory as a basis \cite{Watson:2006yq,Watson:2007mz}.  We derive 
the relevant \DS equations (or their modification from the pure Yang--Mills 
theory) for the two-point functions and then consider their perturbative 
counterparts.  Using techniques based on integration by parts and 
differential equations, we evaluate the noncovariant integrals and present 
the results for the one-loop, two-point dressing functions. 

The paper is organized as follows.  We begin in the next section by 
considering the first order formalism and deriving the relevant field 
equations of motion.  The Feynman rules and general decompositions of the 
two-point functions are obtained in Section~3.  In Section~4, the \DS 
equations are presented in detail.  The one-loop perturbative dressing 
functions are derived in Section~5.  The evaluation of the massive 
noncovariant integrals arising in the one-loop calculations is presented in 
Section~6. The results for the two-point functions are presented in 
Section~7.  A summary and an outlook are given in Section~8. Various 
technical details are discussed in the Appendices.

%-----------------------------------------------------------------------------
\section{Functional formalism}
\setcounter{equation}{0}

Throughout this work, we will use the notations and conventions introduced 
in \cite{Watson:2006yq,Watson:2007mz}.  We work in
Minkowski space (until the perturbative integrals are to be explicitly 
evaluated whereupon we analytically continue to Euclidean space), with the 
metric $g_{\mu\nu}=\textrm{diag}(1, -\vec 1)$.  Greek letters ($\mu,\nu...$) 
denote Lorentz indices, roman letters ($i,j...$) denote spatial indices and 
superscripts ($a,b,c...$) stand for color indices in the adjoint 
representation.  Also, configuration space coordinates may be denoted with 
subscript ($x,y,z...$) when no confusion arises.  The Dirac $\ga$ matrices 
satisfy $\{\ga^\mu,\ga^\nu\}=2g^{\mu\nu}$.  The Yang--Mills and quark 
actions are given by:
\bea
\cs_{\mathit{YM}}&=&\int d^4x\left[-\frac14 F^a_{\mu\nu}F^{a\mu\nu}\right],\\
\cs_{q}&=&\int d^4x\textrm{~}\bar q(\imath\gamma^{0}D_{0}
+\imath \gamma^{j}D_{j}-m)q
\label{eq:qld}
\eea
and the QCD action, $\cs_{QCD}=\cs_{YM}+\cs_{q}$, is invariant under a local 
$SU(N_c)$ gauge transform.  In the above quark action, $q$ ($\ov{q}$) 
denotes the (conjugate) quark field, the Dirac and color indices in 
fundamental representation are implicit and we have $N_f$ flavors of 
identical quarks and $N_c$ colors. The notation $\gamma^i$ refers to the 
spatial component of the Dirac $\gamma$ matrices, where the minus sign 
arising from the metric has been explicitly extracted when appropriate.  The 
temporal and spatial components of the covariant derivative (also implicitly 
in the fundamental color representation) are given by:
\bea
D^{0}&=&\pd^{0}-\imath g T^{c}\si^{c},\nonumber\\
D_i&=&\pd_i+\imath g T^{c} A^{c}_i,
\label{eq:covder}
\eea
where we also rename the temporal component of the gauge field ($A^{0a}$) 
to $\si^a$.  The covariant derivative and (in the Yang--Mills action) the 
field strength tensor $F$ are defined in terms of the gauge potential $A$:
\be
F^a_{\mu\nu}=\pd_{\mu}A_{\nu}^a-\pd_{\nu}A_{\mu}^a
+g f^{abc}A_{\mu}^bA_{\nu}^c,
\ee
where the $f^{abc}$ are the (totally antisymmetric) structure constants of 
the $SU(N_c)$ group, whose (hermitian) generators satisfy 
$[T^a,T^b]=\imath f^{abc}T^c$ and we use the normalization condition 
$\mathrm{tr} (T^aT^b)=\de^{ab}/2$.  For later use, the color factor (Casimir 
invariant) for the gap equation will be written $C_F=(N_c^2-1)/2N_c$.

Consider now the functional integral:
\be
Z=\int\cd\Phi\exp{\left\{\imath\cs_{QCD}\right\}},
\label{eq:funcint}
\ee
where $\cd\Phi$ denotes the functional integration measure for all fields.  
Since the action is invariant under gauge transformations, $Z$ is divergent 
by virtue of the integration over the gauge group. To overcome this problem 
we use the Faddeev-Popov technique and introduce a gauge-fixing term (for 
Coulomb gauge: $\div\cdot\vec{A}^a=0$) along with an associated ghost term 
in standard fashion.  At this stage, we also convert to the first order 
formalism.  This entails introducing three auxiliary fields: 
$\vec{\pi}$, $\phi$ and $\tau$ in order to linearize the action with respect 
to the temporal component of the gauge field ($\si$) in local fashion.  
Classically, $\vec{\pi}$ would be the momentum conjugate to $\vec{A}$ and is 
here transverse ($\div\cdot\vec{\pi}^a=0$); $\phi$ is a scalar field  such 
that $-\div\phi^a$ is the longitudinal component of the conjugate momentum 
field and $\tau$ is a trivial Lagrange multiplier field.  All this is 
described in detail in Ref.~\cite{Watson:2006yq} and is not repeated here.  
Indeed, in this work the details of the gauge-fixing and first order 
formalism are unimportant because the quarks do not directly couple to any 
of these fields (including the ghosts).  What is however important later on 
is that these extra fields will formally enter the discussion of the 
Legendre transform (through partial functional derivatives) which, in 
principle, gives additional terms but which vanish at one-loop order 
perturbatively.  Thus, the reader need only be aware of the existence of 
such fields and be assured that the specific details are not relevant to 
the present study.

The full generating functional of the theory is given by the functional 
integral, \eq{eq:funcint}, in the presence of sources.  Making the sources 
relevant to this work explicit and denoting the rest with dots, we have:
\be
Z[J]=\int\cd\Phi\exp{\left\{
\imath\cs_{QCD}+
\imath \int d^4 x(\bar\chi_xq_x+\bar q_x\chi_x+\rho^a\si^a
+\s{\vec{J}^a}{\vec{A}^a}+\ka^a\phi^a
+\s{\vec{K}^a}{\vec{\pi}^a}+\ldots)
\right\}}.
\label{eq:Z}
\ee
The field equations of motion (from which the \DS equations follow) are 
derived from the observation that the integral of a total derivative 
vanishes, up to possible  boundary terms. We use the usual assumption that 
these boundary terms do not contribute \cite{Watson:2006yq}.  For the quark 
field (we will return to other fields below), we have:
\be
\int\cd\Phi\frac{\delta}{\delta\imath \bar q_{x}}
\exp
{\left\{
\imath\cs_{YM}
+\imath \int d^4 x
\left[\bar q_{x}(\imath\ga^{0}D_{0 x}+\imath \ga^{j}D_{jx}-m )q_{x}
+\bar\chi_{x}q_{x}+\bar q_{x}\chi_{x}\right]
+\ldots\right\}}
=0
\label{eq:gen3}
\ee
(again the dots represent those source terms that do not play a role here).  
Using the expression for the components of the covariant derivative, 
\eq{eq:covder}, it follows that
\be
\int\cd\Phi
[(\imath \gamma^{0}\pd_{0x}+\imath\gamma^{k}\nabla_{kx}
+ gT^{c}\gamma^{0}\si_{x}^{c}- gT^{c}\gamma^{k} A_{kx}^{c}-m)q_{x}+\chi_{x}]
\exp{\left\{\imath\cs\right\}}=0,
\label{eq:gen5}
\ee
where $\cs$ is the full action plus source terms.

The generating functional, $Z[J]$, generates both connected and 
disconnected Green's functions. However, in practice we work with connected 
two-point and one-particle irreducible $n$-point  Green's functions.  The 
generating functional of connected Green's functions is $W[J]$, where 
$Z=e^{W}$.  We introduce a bracket notation for the functional derivatives 
of $W$, such that for a generic source denoted by $J_\al$ (the index denotes 
both the type and all other possible attributes of the field):
\be
\frac{\delta W}{\delta \imath J_{\alpha}}= \ev{\imath J_{\alpha}}.
\ee
Converting \eq{eq:gen5} into derivatives of $W[J]$ we obtain :
\bea
(\imath\gamma^{0}\pd_{0x}
+\imath\gamma^{k}\nabla_{kx}-m)\ev{\imath\bar\chi_{x}}
+g T^{c}\left\{
\gamma^{0}[\ev{\imath\rho_{x}^{c}}\ev{\imath\bar\chi_{x}}
+\ev{\imath\rho_{x}^{c}\imath\bar\chi_{x}}]
-\gamma^{k}[\ev{\imath J_{kx}^{c}}\ev{\imath\bar\chi_{x}}
+\ev{\imath J_{kx}^{c}\imath\bar\chi_{x}}]\right\}
+\chi_{x}
=0.\nonumber\\
\label{eq:eomW}
\eea
We define the generic classical field (we use the same notation for the 
classical fields and for the quantum fields which are integrated over since 
no confusion will arise) to be:
\be
\Phi_{\al}=\frac1Z\int
\cd\Phi\Phi_{\al}\exp{\left\{\imath\cs\right\}}=
\frac1Z\frac{\delta Z}{\delta\imath J_{\alpha}}.
\ee
The generating functional of the proper (one-particle irreducible) Green's 
functions is the effective action, $\G[\Phi]$, (which is a function of the 
classical fields) and is defined via the Legendre transform of $W[J]$:
\be
\G[\Phi]=W[J]-\imath \Phi_{\alpha} J_{\alpha}.
\label{eq:lt}
\ee
(We use the common convention that summation over all discrete indices and 
integration over continuous arguments is implicit).   This gives:
\bea 
\ev{\imath J_{\alpha}}=\frac{\delta W}{\delta \imath J_\al}=\Phi_{\alpha}
\textrm{~~~and~~~}
\ev{\imath\Phi_{\alpha}}=\frac{\delta\G}{\delta\imath\Phi_\al}=- J_{\alpha}.
\eea
We have used the same bracket notation to denote derivatives of $\G$ with 
respect to fields -- no confusion arises since we never mix derivatives 
with respect to sources and fields.  Note that care must be taken to 
observe the correct minus signs associated with the Grassmann fields and 
sources.  The equation of motion, \eq{eq:eomW}, in terms of proper 
functions (and from which we will derive the quark gap equation) now reads:
\bea
\ev{\imath \bar q_{x}} =
-\imath(\imath\gamma^{0}\pd_{0 x}+\imath\gamma^{k}\nabla_{kx}-m )\imath q_{x}
+g T^{c}\gamma^{0}\left[\si_{x}^{c}q_{x}
+\ev{\imath\rho_{x}^{c}\imath\bar\chi_{x}}\right]
-g T^{c}\gamma^{k}\left[A_{kx}^{c}q_{x}
+\ev{i J_{kx}^{c}\imath\bar\chi_{x}}\right].
\label{eq:qeom}
\eea
In a similar fashion, the quark contributions to the field equations of 
motion for $\ev{\imath\si_x^a}$ and $\ev{\imath A_{ix}^a}$ are given by 
(the rest of these equations are simply the Yang--Mills expressions derived 
previously in Ref.~\cite{Watson:2006yq} and are not important here):
\bea
\ev{\imath\si_x^a}&=&
g\bar{q}_{x}T^a\gamma^{0}q_{x}
+g\mathrm{Tr}
\left\{T^a\gamma^{0}\ev{\imath\bar\chi_{x}\imath\chi_{x}}\right\}+\ldots~,
\label{eq:sidse0}\\
\ev{\imath A_{ix}^a}&=&
-g\bar{q}_xT^a\gamma^{i}q_x
-g\mathrm{Tr}
\left\{T^a\gamma^{i}\ev{\imath\bar\chi_{x}\imath\chi_{x}}\right\}+\ldots~,
\label{eq:adse0}
\eea
where the trace is over Dirac and fundamental color indices.

%-----------------------------------------------------------------------------
\section{Feynman rules and decompositions}
\setcounter{equation}{0} 

Let us now discuss the Feynman rules and decompositions of the Green's 
functions.  The tree-level quark propagator can be derived directly from 
the quark equation of motion in terms of connected functions, \eq{eq:eomW}, 
by functionally differentiating and neglecting the interaction terms. For 
the noncovariant case here we obtain (after Fourier transforming to momentum 
space):
 \be 
W_{\bar qq}^{(0)}(k)=
-\imath \frac{\gamma^0 k_0-\gamma^i k_i+m}{k_0^2-\vec k^2-m^2}.
\label{eq:treelevelquarkprop}
\ee
Later on (where appropriate), we will use the usual notation 
$\kslash=\gamma^0 k_0-\gamma^i k_i$.  The tree-level gluon propagators 
needed in this work have been derived in \cite{Watson:2006yq} and are given 
by:
\bea
&&W_{A A ij}^{(0)}(k)=\frac{\imath t_{ij}(k)}{k_{0}^2-\vec{k}^2},
\textrm{~~~}W_{\si\si}^{(0)}(k)=\frac{\imath }{\vec{k}^2}
\label{eq:treelevelgluonprop}
\eea
where $t_{ij}(k)=\delta_{ij}-k_ik_j/\vec{k}^2$ is the transverse spatial 
projector.  It is understood that the denominator factors involving both 
temporal and spatial components implicitly carry the Feynman 
prescription, i.e.,
\be
\frac{1}{(k_0^2-\vec k^2)}\rightarrow\frac{1}{(k_0^2-\vec k^2+\imath 0_+)},
\ee 
such that the analytic continuation to the Euclidean space can be performed.

The tree-level quark proper two-point function is derived from the quark 
equation of motion in terms of proper functions, \eq{eq:qeom}:
\be
\G_{\bar qq}^{(0)}(k)=\imath (\kslash-m).
\ee
There are two tree-level quark-gluon vertices, spatial and temporal, again 
obtained by taking the appropriate functional derivatives:
\bea
\G_{\bar q q\si}^{(0)a}&=&gT^{a}\ga^{0},\nonumber\\
\G_{\bar q qAj}^{(0)a}&=&-gT^{a}\ga^{j}.
\label{eq:treelevelquarkvertex}
\eea

In addition to their tree-level forms, we will also require the general 
decompositions for the quark two-point functions (connected and proper) and 
the relationship between them.  Because we work in a noncovariant setting, 
the usual arguments must be modified to include separately the temporal and 
spatial components.  Starting with the quark propagator, we observe that
\be
W_{\ov{q}q}(k^0,\vec{k})
\sim\int\cd\Phi\,\ov{q}q\,\exp{\left\{\imath\cs\right\}}
\ee
such that under both time-reversal and parity transforms, the propagator 
will remain unchanged -- the bilinear combination of fields is scalar.  
Since the propagator depends on both $k^0$ and $\vec{k}$, it has thus 
\emph{four} components in distinction to the covariant case where there 
are only two.  We thus write
\be
W_{\bar qq}(k)=-\frac{\imath}{k_0^2-\vec k^2-m^2}
\left\{ k_0\ga^{0}F_t(k)-k_i\ga^{i}F_s(k)+M(k)+k_0k_i\ga^0\ga^iF_d(k)
\right\}
\ee
where all dressing functions are functions of both $k_0^2$ and $\vec{k}^2$.  
At tree-level, $F_t=F_s=1$, $F_d=0$ and $M=m$.  The last term with $F_d$ 
has no covariant counterpart and in fact will only appear (if at all) at 
two-loop order and beyond, as will be justified below.  For the proper 
two-point function, the same arguments apply and we write
\be
\G_{\bar qq}(k)=\imath
\left\{k_0\ga^{0}A_t(k)-k_i\ga^{i}A_s(k)-B_m(k)+k_0k_i\ga^0\ga^iA_d(k)\right\}
\label{eq:gap_gendecom}
\ee
and we will refer to $A_t, A_s$ and $B_m$ as the temporal, spatial and 
massive components, respectively.  Again the last component ($A_d$) has 
no covariant counterpart and will only appear at two-loops or beyond.  The 
relationship between the connected and proper two-point functions is 
supplied via the Legendre transform in standard fashion and we have
\be
\G_{\bar qq}(k)W_{\bar qq}(k)=1.
\ee
In terms of the dressing functions, this gives
\bea
F_t&=&\frac{(k_0^2-\vec k^2-m^2)A_t}
{k_0^2A_t^2-\vec k^2A_s^2-B_m^2+k_0^2\vec{k}^2A_d^2},\nonumber\\
F_s&=&\frac{(k_0^2-\vec k^2-m^2)A_s}
{k_0^2A_t^2-\vec k^2A_s^2-B_m^2+k_0^2\vec{k}^2A_d^2},\nonumber\\
M&=&\frac{(k_0^2-\vec k^2-m^2)B_m}
{k_0^2A_t^2-\vec k^2A_s^2-B_m^2+k_0^2\vec{k}^2A_d^2},\nonumber\\
F_d&=&\frac{(k_0^2-\vec k^2-m^2)A_d}
{k_0^2A_t^2-\vec k^2A_s^2-B_m^2+k_0^2\vec{k}^2A_d^2}.
\label{eq:fullwd}
\eea
Let us now discuss possible appearance of the genuinely noncovariant term 
corresponding to the dressing function $A_d$.  In this work it will arise, 
if at all, from the one-loop perturbative form of the quark self-energy 
(see later for details).  For now, we can anticipate that since the 
tree-level quark propagator does not contain a term with $k_0k_i\ga^0\ga^i$ 
and in the self-energy loop with two tree-level vertices we only have 
either two $\ga^0$ or two $\ga^i$ matrices together (the gluon propagator 
is either purely temporal or spatial), then there is no one-loop 
contribution that has the overall structure $\ga^0\ga^i$.  This means that 
$A_d=0$ at one-loop order perturbatively.  For the rest of the dressing 
functions, we then get the simplified set of relations:
\bea
F_{t}=\frac{(k_0^2-\vec k^2-m^2)A_t}
{k_0^2A_t^2-\vec k^2A_s^2-B_m^2},\textrm{~~~}
F_{s}=\frac{(k_0^2-\vec k^2-m^2)A_s}
{k_0^2A_t^2-\vec k^2A_s^2-B_m^2},\textrm{~~~}
M=\frac{(k_0^2-\vec k^2-m^2)B_m}{k_0^2A_t^2-\vec k^2A_s^2-B_m^2}.
\label{eq:b}
\eea
We emphasize that these relations only hold up to one-loop perturbatively 
--- in possible future studies it must be recognized that the fourth Dirac 
structure, $\gamma^0\gamma^i$, may enter in a nontrivial fashion and that 
the set of relations \eq{eq:fullwd} should be used.

Evaluation of the quark contributions to the $W_{AA}$ and $W_{\si\si}$ 
propagators is a direct extension of the results already obtained in 
\cite{Watson:2006yq} for the Yang--Mills sector.

%-----------------------------------------------------------------------------
\section{Derivation of the Dyson--Schwinger equations}
\setcounter{equation}{0}

We start the derivation of the gap equation by taking the functional 
derivative of the quark field equation of motion (in configuration space), 
\eq{eq:qeom}, with respect to $\imath q_{w}$:
\be
\label{eq:qq}
\ev{\imath\bar q_{x}\imath q_{w}} =
\imath(\imath\gamma^{0}\pd_{0 x} +\imath\gamma^{k}\nabla_{kx}-m)\delta (x-w)
-\int d^4yd^4z\,\de(x-y)\de(x-z)\left[\G_{\bar qq\si}^{(0)a}
\frac{\delta}{\delta\imath q_{w}}\ev{\imath\rho_{y}^{a}\imath\bar\chi_{z}}
+\G_{\bar qq Aj}^{(0)a}\frac{\delta}{\delta\imath q_{w}}
\ev{\imath J_{jy}^{a}\imath\bar\chi_{z}}\right].
\ee
In the above we have used the configuration space definitions of the 
tree-level quark-gluon vertices (extracting the trivial $\de$-function 
dependence in configuration space) and omit those terms which will 
eventually vanish when the sources are set to zero.
%-----------------------------------------------------------------------------
\begin{figure}[t]
\vspace{0.2cm}
\includegraphics[width=0.5\linewidth]{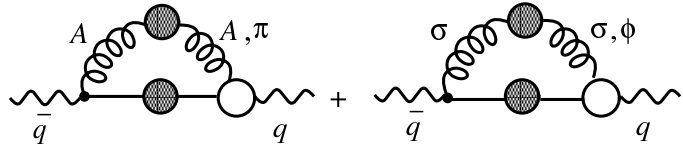}
\caption{\label{fig:gapeq}Full nonperturbative diagram for the quark 
self-energy. Filled circles denote dressed propagators and empty circles 
denote dressed vertices. Springs denote connected (propagator) functions, 
solid lines denote quark propagators and wavy lines denote the external legs 
of the proper functions.}
\end{figure}
%-----------------------------------------------------------------------------
We use partial differentiation to calculate the terms in the bracket.  
Notice that simply because of the presence of the $\vec{\pi}$ and $\phi$ 
fields arising in the first order formalism, we must allow for the 
additional terms so generated (these terms will vanish when we consider 
the one-loop perturbative case):
\bea
\frac{\delta}{\delta\imath q_{w}}\ev{\imath\rho_{y}^{a}\imath\bar\chi_{z}}&=&
-\int d^4vd^4u
\ev{\imath\bar\chi_{z}\imath\chi_{v}}\ev{\imath\rho_{y}^{a}\imath\rho_{u}^{b}}
\ev{\imath \bar q_{v}\imath q_{w}\imath \si_{u}^{b}}\nonumber\\
&&-\int d^4vd^4u
\ev{\imath\bar\chi_z\imath\chi_v}\ev{\imath\rho_y^a\imath\kappa_{u}^{b}}
\ev{\imath \bar q_{v}\imath q_{w}\imath \phi_{u}^{b}},
\label{eq:rhoqq}\\
\frac{\delta}{\delta\imath q_{w}}\ev{\imath J_{jy}^{a}\imath\bar\chi_{z}}&=&
-\int d^4vd^4u
\ev{\imath\bar\chi_{z}\imath\chi_{v}}\ev{\imath J_{yj}^{a}\imath J_{uk}^{b}}
\ev{\imath \bar q_{v}\imath q_{w}\imath A_{uk}^b}\nonumber\\
&&-\int d^4vd^4u
\ev{\imath\bar\chi_{z}\imath\chi_{v}}\ev{\imath J_{yj}^{a}\imath K_{uk}^{b}}
\ev{\imath \bar q_{v}\imath q_{w}\imath \pi_{uk}^{b}}.
\label{eq:jqq}
\eea
Inserting the above expressions into \eq{eq:qq} and Fourier transforming 
into momentum space we obtain the quark \DS (or gap) equation
\bea
\G_{\bar q q}(k)&=&
\imath(\gamma^{0}k_{0} -\gamma^{j}k_{j}-m)
+\int \dk{\w}
\G_{\bar q q\si}^{(0)a}(k,-\w,\w -k)W_{\bar q q}(\w)
\G_{\bar q q\si}^{b}(\w,-k,k-\w)W_{\si\si}^{ab}(k-\w)
\nonumber\\
&&+\int \dk{\w}
\G_{\bar q q\si}^{(0)a}(k,-\w,\w -k)W_{\bar q q}(\w)
\G_{\bar q q\phi}^{b}(\w,-k,k-\w)W_{\si\phi}^{ab}(k-\w)
\nonumber\\
&&+\int\dk{\w}
\G_{\bar q qAi} ^{(0)a}(k,-\w,\w -k)W_{\bar q q}(\w)
\G_{\bar qqAj}^{b}(\w,-k,k-\w)W_{AAij}^{ab}(k-\w)
\nonumber\\
&&+\int\dk{\w}
\G_{\bar q qAi} ^{(0)a}(k,-\w,\w -k)W_{\bar q q}(\w)
\G_{\bar qq\pi j}^{b}(\w,-k,k-\w)W_{A\pi ij}^{ab}(k-\w),
\label{eq:gapeqms}
\eea
where $\dk{\w}=d^4 \w/(2\pi)^{4}$.  The self-energy corrections are 
presented diagrammatically in \fig{fig:gapeq}.  We see that the $\pi$ and 
$\phi$ fields do make a contribution thanks to the existence of mixed 
propagators in the first order formalism.  But, as emphasized, these 
contributions will drop out at one-loop order perturbatively because of 
the absence of corresponding tree-level vertices, i.e., that there exist 
no direct interaction terms in the action between the quark fields and the 
auxiliary fields of the first order formalism.

\begin{figure}[t]
\vspace{0.2cm}
\includegraphics[width=0.2\linewidth]{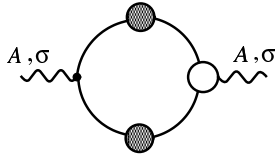}
\caption{\label{fig:Asiqq}One-loop diagram for the quark contributions to 
the gluon proper two-point functions.  Filled circles denote dressed 
propagators and empty circles denote dressed vertices.  Solid lines denote 
quark propagators and wavy lines denote the external legs of the proper 
functions.}
\end{figure}

We next consider the quark contributions to the proper two-point functions 
$\G_{\si\si}$, $\G_{\si A}$ and $\G_{AA}$.  Starting with the $\si$ equation 
of motion \eq{eq:sidse0} and following the same procedure as for the gap 
equation we derive the quark contribution to the proper two-point function 
$\G_{\si\si}$ in configuration space (trace over Dirac and fundamental 
color indices):
\bea
\ev{\imath\si_x^a\imath\si_w^b}_{(q)}&=&
-\mathrm{Tr}\int d^4yd^4zd^4ud^4v
\G_{\bar q q\si}^{(0)a}(z,y,x)
\ev{\imath\bar \chi_{y}\imath \chi_{u}}
\ev{\imath \bar q_{u}\imath q_{v}\imath\si_{w}^{b}}
\ev{\imath \bar\chi_{v}\imath \chi_{z}}.
\label{eq:qqsigma_mom}
\eea
Taking the Fourier transform, we get in momentum space:
\bea
\G_{\si\si(q)}^{ab}(k)=
-\mathrm{Tr}\int\dk{\w}
\G_{\bar qq\si}^{(0)a}(\w-k,-\w,k)W_{\bar qq}(\w)
\G_{\bar qq\si }^{b}(\w,k-\w,-k)W_{\bar qq }(\w-k).
\label{eq:siqdse}
\eea
Similarly we obtain:
\bea
\G_{\si Ai(q)}^{ab}(k)&=&-\mathrm{Tr}\int\dk{\w}
\G_{\bar qq\si}^{(0)a}(\w-k,-\w,k)W_{\bar qq}(\w)
\G_{\bar qqAi}^{b}(\w,k-\w,-k)W_{\bar qq }(\w-k),
\label{eq:Aqdse}\\
\G_{AAij(q)}^{ab}(k)&=&-\mathrm{Tr}\int\dk{\w}
\G_{\bar qqAi}^{(0)a}(\w-k,-\w,k)W_{\bar qq}(\w)
\G_{\bar qq Aj}^{b}(\w,k-\w,-k)W_{\bar qq }(\w-k).
\label{eq:siAqdse}
\eea
These loop contributions are shown collectively in Fig.~\ref{fig:Asiqq}.

%-----------------------------------------------------------------------------
%-----------------------------------------------------------------------------

%-----------------------------------------------------------------------------
\section{One-loop perturbative two-point functions}
\setcounter{equation}{0}

Let us now consider the one-loop perturbative expansions of the two-point 
functions derived in the previous section.  Although so far the formalism 
has been presented in 4-dimensional Minkowski space, in order to evaluate 
the resulting loop integrals we have to make the analytic continuation to 
Euclidean space ($k_0\rightarrow \imath k_4$), where we denote the 
temporal component of the Euclidean 4-momentum $k_4$ such that 
$k^2=k_4^2+\vec k^2$.  Additionally, to regularize the integrals, 
dimensional regularization is employed with the Euclidean space integration 
measure
\be
\dk{\w}=\frac{d\w_4 d^d \vec\w}{(2\pi)^{d+1}}
\ee
where $d=3-2\e$ is the spatial dimension.  To preserve the dimension of the 
action we must assign a dimension to the coupling through the replacement
\be
g^2\rightarrow g^2\mu^\e,
\ee
where $\mu$ is the square of a non-vanishing mass scale (which may be later 
associated with a renormalization scale).  The perturbative expansion of 
the two-point dressing function is generically written as:
\be
\G=\G^{(0)}+g^2\G^{(1)}
\ee 
where the factor $\mu^{\e}$ is included in $\G^{(1)}$ such that the new 
coupling and $\G^{(1)}$ are dimensionless.

In the full nonperturbative gap equation, \eq{eq:gapeqms}, we first insert 
the various tree-level vertices and propagators given by 
Eqs.~(\ref{eq:treelevelquarkprop}), (\ref{eq:treelevelgluonprop}) and 
(\ref{eq:treelevelquarkvertex}).  Then we insert the general decomposition 
of the proper two-point functions, \eq{eq:gap_gendecom}, occurring on the 
left-hand side of the gap equation, take the Dirac projection, solve the 
color and tensor algebra and lastly perform the Wick rotation.  The 
one-loop temporal, spatial and massive components of the quark gap equation 
in Euclidean space read (recall that $C_F=(N_c^2-1)/2N_c$ ):
\bea
A_t(k)&
=&1-g^2\mu^\e C_F\frac{1}{k_4^2}\int\dk{\w}
\left\{\frac{k_4\w_4}{(\w^2+m^2)(\vec k-\vec\w)^2}
-\frac{k_4\w_4(\textrm{d}-1)}{(\w^2+m^2)(k-\w)^2}\right\},
\label{eq:atmassive}\\
A_s(k)&
=&1-g^2\mu^\e C_F\frac{1}{\vec k^2}
\int\dk{\w}\left\{
-\frac{2[\vec k\cdot(\vec k-\vec\w)][\vec\w\cdot(\vec k-\vec\w)]}
{(\w^2+m^2)(k-\w)^2(\vec k-\vec\w)^2}
-\frac{\vec k\cdot\vec\w}{(\w^2+m^2)(\vec k-\vec\w)^2}
+\frac{\vec k\cdot\vec\w(3-\textrm{d})}{(\w^2+m^2)(k-\w)^2}
\right\},\nonumber\\
\label{eq:asmassive}\\
B_m(k)&=&m+m g^2\mu^\e C_F\int\dk{\w}
\left\{\frac{1}{(\w^2+m^2)(\vec
 k-\vec\w)^2}+\frac{(\textrm{d}-1)}{(\w^2+m^2)(k-\w)^2}\right\}.
\label{eq:bmassive}
\eea
As mentioned earlier, the possible contribution corresponding to the 
genuinely noncovariant dressing function $A_d$ does not appear at 
one-loop.  To evaluate the integrals occurring in \eq{eq:asmassive}, it 
is helpful to use the identity:
\bea
\s{\vec{k}}{\vec{\w}}=\frac12\left[k^2+\w^2-(k-\w)^2\right]-k_4\w_4,
\eea
which enables us to rewrite $A_s$ as a combination of more straightforward 
integrals:
\bea
A_s(k)&=&1-g^2\mu^\e C_F\frac{1}{\vec k^2}
\int\dk{\w}\left\{
-\frac12\frac{(k^2+m^2)^2}{\w^2[(k-\w)^2+m^2]\vec\w^2}
+\frac{2(k^2+m^2)k_4\w_4}{\w^2[(k-\w)^2+m^2]\vec\w^2}
+\frac{2\e \vec k^2+2k_4^2}{[(k-\w)^2+m^2]\w^2}\right.\nonumber\\
&&\left.+\frac{2\vec k\cdot\vec\w  (1-\e)}{[(k-\w)^2+m^2]\w^2}
-\frac12 \frac{m^2+3 k^2}{[(k-\w)^2+m^2]\vec\w^2}
\right\}.
\label{eq:asmassive1}
\eea

To determine the quark contributions to the various proper two-point gluon 
dressing functions given by Eqs. (\ref{eq:siqdse}-\ref{eq:siAqdse}) 
(presented in \fig{fig:Asiqq}), we again insert the tree-level factors given 
by Eqs.~(\ref{eq:treelevelquarkprop}),(\ref{eq:treelevelgluonprop}) and 
(\ref{eq:treelevelquarkvertex}), solve the color and tensor algebra and 
perform a Wick rotation.  The one-loop integral expressions are:
\bea
&&\vec k^2\G_{\si\si(q)}^{(1)}(k)=\mu^\e N_{f}2
\int\dk{\w}\frac{\vec\w^2-\w_4^2-\vec\w\cdot\vec k+\w_4k_4+m^2}
{(\w^2+m^2)[(k-\w)^2+m^2]},
\label{eq:Gsisi}\\
&&k_i k_4\G_{\si A(q)}^{(1)}(k)=\mu^\e N_{f}4
\int\dk{\w}\frac{\w_i\w_4-k_i\w_4}{(\w^2+m^2)[(k-\w)^2+m^2]},
\label{eq:GsiA}\\
&&\vec k^2t_{ij}(\vec k)\G_{AA(q)}^{(1)}(k)+ k_ik_j\bar\G_{AA(q)}^{(1)}(k)
= 2N_f\mu^\e
\int\dk{\w}\frac{2\w_i\w_j-2\w_ik_j+\delta_{ij}
(\w_4k_4+\vec\w\cdot\vec k-\w_4^2-\vec\w^2-m^2)}{(\w^2+m^2)[(k-\w)^2+m^2]},
\nonumber\\\label{eq:GAA}
\eea
where $\G_{AA(q)}^{(1)}$ and $\bar\G_{AA(q)}^{(1)}$ are the transversal and 
longitudinal components of the proper two-point function 
$\G_{AAij(q)}^{(1)ab}$ (given by \eq{eq:siAqdse}), respectively (see 
Ref.~\cite{Watson:2006yq} for details of decomposition).

%-----------------------------------------------------------------------------
\section{Noncovariant Massive Loop Integrals}
\setcounter{equation}{0}

In the one-loop expansions from the previous section, there are two types 
of integrals arising: those which can be solved using standard techniques 
(such as Schwinger parametrization, Mellin representation -- for details, 
see Appendix~\ref{app:stdint}), and those which require a more esoteric 
approach.  In this section we study the latter variety using a technique 
based on differential equations and integration by parts developed 
previously \cite{Watson:2007mz}.  We will consider the two integrals:
\bea
A_m(k_4^2,\vec{k}^2)&=&\int\frac{\dk{\w}}{\w^2[(k-\w)^2+m^2]\vec{\w}^2},
\label{eq:adef}\\
A^4_m(k_4^2,\vec{k}^2)&=&
\int\frac{\dk{\w}\,\w_4}{\w^2[(k-\w)^2+m^2]\vec{\w}^2}.
\label{eq:a4def}
\eea

%----------------------------------------------------------------------------
\subsection{Derivation of the differential equations}

Let us first write Eqs.~(\ref{eq:adef}) and (\ref{eq:a4def}) in the general 
form ($n=0,1$) 
\be
I^n(k_4^2,\vec{k}^2)=\int\frac{\dk{\w}\,\w_4^n}{\w^2[(k-\w)^2+m^2]\vec{\w}^2}.
\label{eq:idef}
\ee
In this derivation  $k_4^2$ and $\vec{k}^2$ are treated as variables 
whereas the mass, $m$, is treated as a parameter.  The two first 
derivatives are:
\bea
k_4\frac{\pd I^n}{\pd k_4}&=&\int\frac{\dk{\w}\,\w_4^n}{\w^2[(k-\w)^2+m^2]
\vec{\w}^2}\left\{-2\frac{k_4(k_4-\w_4)}{(k-\w)^2+m^2}\right\},
\label{eq:ade0}\\
k_k\frac{\pd I^n}{\pd k_k}&=&\int\frac{\dk{\w}\,\w_4^n}{\w^2[(k-\w)^2+m^2]
\vec{\w}^2}
\left\{-2\frac{\s{\vec{k}}{(\vec{k}-\vec{\w})}}{(k-\w)^2+m^2}\right\}.
\label{eq:adevec}
\eea
There are also two integration by parts identities:
\bea
0=\int\dk{\w}
\frac{\pd}{\pd\w_4}\frac{\w_4^{n+1}}{\w^2[(k-\w)^2+m^2]\vec{\w}^2}
&=&\int\frac{\dk{\w}\,\w_4^n}{\w^2[(k-\w)^2+m^2]\vec{\w}^2}
\left\{n+1-2\frac{\w_4^2}{\w^2}-2\frac{\w_4(\w_4-k_4)}{(k-\w)^2+m^2}\right\},
\label{eq:ibpA}\\
0=\int\dk{\w}
\frac{\pd}{\pd\w_i}\frac{\w_i\w_4^n}{\w^2[(k-\w)^2+m^2]\vec{\w}^2}
&=&\int\frac{\dk{\w}\,\w_4^n}{\w^2[(k-\w)^2+m^2]\vec{\w}^2}
\left\{d-2-2\frac{\vec{\w}^2}{\w^2}-2\frac{\s{\vec{\w}}{(\vec{\w}-\vec{k})}}
{(k-\w)^2+m^2}\right\}.
\label{eq:ibpA4}
\eea
Adding these two expressions gives
\be
0=\int\frac{\dk{\w}\,\w_4^n}{\w^2[(k-\w)^2+m^2]\vec{\w}^2}
\left\{d+n-3-2\frac{\w\cdot(\w-k)}{(k-\w)^2+m^2}\right\}.
\label{eq:ibpSum}
\ee
Expanding the numerator factor, we can rewrite \eq{eq:ibpSum} as
\be
0=\int\frac{\dk{\w}\,\w_4^n}{\w^2[(k-\w)^2+m^2]\vec{\w}^2}
\left\{d+n-4+\frac{k^2-\w^2+m^2}{(k-\w)^2+m^2}\right\}.
\label{eq:ibpSum1}
\ee
Combining this with \eq{eq:ibpA} then yields:
\bea
&&\int\frac{\dk{\w}\,\w_4^n}{\w^2[(k-\w)^2+m^2]\vec{\w}^2}
\left\{-2\frac{k_4(k_4-\w_4)}{(k-\w)^2+m^2}\right\}=
\int\frac{\dk{\w}\,\w_4^n}{\w^2[(k-\w)^2+m^2]\vec{\w}^2}
\left[\frac{2 k_4^2}{k^2+m^2} (n+d-4)-n+1\right]\nonumber\\
&&+\frac{\vec{k}^2+m^2}{k^2+m^2}
\int\frac{\dk{\w} \w_4^n}{[(k-\w)^2+m^2]^2\vec{\w}^2}
-2\int\frac{\dk{\w}\w_4^n}{\w^4[(k-\w)^2+m^2]}
-2\int\frac{\dk{\w}\w_4^n}{\w^2[(k-\w)^2+m^2]^2}.
\eea
This leads to the temporal differential equations for $A_m$ and $A^4_m$:
\bea
k_4\frac{\pd A_m}{\pd k_4}&=&\left[1+2\frac{(d-4)k_4^2}{k^2+m^2}\right]A_m
+2\frac{\vec{k}^2+m^2}{k^2+m^2}\int\frac{\dk{\w}}{[(k-\w)^2+m^2]^2\vec{\w}^2}
-2\int\frac{\dk{\w}}{\w^4[(k-\w)^2+m^2]}\nonumber\\
&-&2\int\frac{\dk{\w}}{\w^2[(k-\w)^2+m^2]^2},\label{eq:detemp1}\\
k_4\frac{\pd A^4_m}{\pd k_4}&=&2\frac{(d-3) k_4^2}{k^2+m^2}A_m^4
+2\frac{\vec{k}^2+m^2}{k^2+m^2}
\int\frac{\dk{\w}\,\w_4}{[(k-\w)^2+m^2]^2\vec{\w}^2}
-2\int\frac{\dk{\w}\,\w_4}{\w^4[(k-\w)^2+m^2]}\nonumber\\
&-&2\int\frac{\dk{\w}\,\w_4}{\w^2[(k-\w)^2+m^2]^2}.
\label{eq:detemp2}
\eea
In the same manner, we derive the differential equations involving the 
spatial components:
\bea
k_i\frac{\pd A_m}{\pd k_i}&=&
\left[2-d+2\frac{(d-4)\vec{k}^2}{k^2+m^2}\right]A_m
-\frac{2\vec{k}^2}{k^2+m^2}\int\frac{\dk{\w}}{[(k-\w)^2+m^2]^2\vec{\w}^2}
+2\int\frac{\dk{\w}}{\w^4[(k-\w)^2+m^2]}\nonumber\\
&+&2\int\frac{\dk{\w}}{\w^2[(k-\w)^2+m^2]^2},\label{eq:desp1}\\
k_i\frac{\pd A^4_m}{\pd k_i}&=&
\left[2-d+2\frac{(d-3)\vec{k}^2}{k^2+m^2}\right]A^4_m
-\frac{2\vec{k}^2}{k^2+m^2}
\int\frac{\dk{\w}\,\w_4}{[(k-\w)^2+m^2]^2\vec{\w}^2}
+2\int\frac{\dk{\w}\,\w_4}{\w^4[(k-\w)^2+m^2]}\nonumber\\
&+&2\int\frac{\dk{\w}\,\w_4}{\w^2[(k-\w)^2+m^2]^2}.
\label{eq:desp2}
\eea
It is in fact possible to write down a mass differential equation, but in 
the light of the method presented here this would not bring any new 
information.  However, this third derivative is important because it 
provides an useful check of our solutions (for a detailed discussion, see 
Appendix~\ref{app:intcheck}).

At this point, let us discuss how the differential equations for the 
massless integrals considered in Ref.~\cite{Watson:2007mz} are regained.  
In the massless limit, there are potential ambiguities arising in the 
integrals appearing in Eqs.~(\ref{eq:detemp1} -\ref{eq:desp2}), because in 
part, the limits $m\rightarrow 0$ and $\e\rightarrow 0$ do not 
interchange.  Let us start by considering the following integral given by 
\eq{eq:smi3} (similar arguments apply to all the integrals appearing in 
the differential equations):
\be
I=\int\frac{\dk{\w}}{\vec{\w}^2[(k-\w)^2+m^2]^2}=
\frac{[m^2]^{-1-\e}}{(4\pi)^{2-\e}}\frac{\G(\frac12-\e) \G(1+\e)}{\G(3/2-\e)}
{}_{2}F_{1}\left(1,1+\e;3/2-\e;-\frac{\vec{k}^2}{m^2}\right).
\label{eq:stan_ex}
\ee
It is useful here to invert the argument of the hypergeometric with the 
help of the formula (see, for instance, Ref.~\cite{abramowitz}):
\bea
{}_{2}F_{1}(a,b;c;t)&=&\frac{\G(c)\G(b-a)}{\G(b)\G(c-a)}(-t)^{-a}
{}_{2}F_{1}\left(a,1-c+a;1-b+a;\frac1t\right)\nonumber\\
&&+\frac{\G(c)\G(a-b)}{\G(a)\G(c-b)}(-t)^{-b}
{}_{2}F_{1}\left(b,1-c+b;1-a+b;\frac1t\right).
\label{eq:hyparginv}
\eea
Then we have:
\bea
I\!&=\!&\frac{1}{\vec k^2}\frac{[m^2]^{-\e}\G(\e)}{(4\pi)^{2-\e}}
{}_{2}F_{1}\left(1,\frac 12+\e;1-\e;-\frac{m^2}{\vec{k}^2}\right)
+\frac{[\vec k^2]^{-1-\e}}{(4\pi)^{2-\e}}
\frac{\G\left(\frac12-\e\right)\G(-\e)\G(1+\e)}{\G\left(\frac12-2\e\right)}
{}_{2}F_{1}\left(1+\e,\frac12+2\e;1+\e;-\frac{m^2}{\vec{k}^2}\right).
\nonumber\\
\label{eq:example_ambiguity}
\eea
In the expression above, the problem of the non-interchangeable limits is 
seen explicitly in the first term.  However, when all the integrals 
occurring in the various differential equations are put together, such 
terms explicitly cancel and only the second term of 
\eq{eq:example_ambiguity} (which leads to the correct massless limit) 
contributes. 

Returning to the differential equations, we evaluate the standard integrals 
in terms of $\e$ (see Appendix \ref{app:stdint}) and with the notation 
$x=k_4^2$, $y=\vec{k}^2$ we obtain for $A_m$:
\bea
2x\frac{\pd A_m}{\pd x}&=&\left[1-2(1+2\e)\frac{x}{x+y+m^2}\right]A_m
+2\frac{[m^2]^{-1-\e}}{(4\pi)^{2-\e}}\left\{
\frac{y+m^2}{x+y+m^2}X
{}_{2}F_1\left(1,1+\e;3/2-\e;-\frac{y}{m^2}\right)\right.\nonumber\\
&-& \left.
\frac{\G(-\e)\G(1+\e)}{\G(2-\e)}
{}_{2}F_1\left(2,1+\e;2-\e;-\frac{x+y}{m^2}\right)
-Y{}_{2}F_1\left(1,1+\e;2-\e;-\frac{x+y}{m^2}\right)\right\},
\label{eq:atde0}\\
2y\frac{\pd A_m}{\pd y}&=&\left[-1+2\e-2(1+2\e)\frac{y}{x+y+m^2}\right]A_m
-2\frac{[m^2]^{-1-\e}}{(4\pi)^{2-\e}}\left\{
\frac{y}{x+y+m^2}X
{}_{2}F_1\left(1,1+\e;3/2-\e;-\frac{y}{m^2}\right)\right.\nonumber\\
&-&\left. 
\frac{\G(-\e)\G(1+\e)}{\G(2-\e)}
{}_{2}F_1\left(2,1+\e;2-\e;-\frac{x+y}{m^2}\right)
-Y {}_{2}F_1\left(1,1+\e;2-\e;-\frac{x+y}{m^2}\right)\right\}
\label{eq:asde0}
\eea
and for the integral $A^4=k_4\ov{A}_m$:
\bea
2x\frac{\pd \ov{A}_m}{\pd x}&=&\left[-1-4\e\frac{x}{x+y+m^2}\right]\ov{A}_m
+2\frac{[m^2]^{-1-\e}}{(4\pi)^{2-\e}}\left\{
\frac{y+m^2}{x+y+m^2}X
{}_{2}F_1\left(1,1+\e;3/2-\e;-\frac{y}{m^2}\right)\right.\nonumber\\
&-&\left. \frac{Y}{2-\e}
\left[{}_{2}F_1\left(2,1+\e;3-\e;-\frac{x+y}{m^2}\right)+(1-\e)
{}_{2}F_1\left(1,1+\e;3-\e;-\frac{x+y}{m^2}\right)
\right]\right\},\label{eq:at4de0}\\
2y\frac{\pd \ov{A}_m}{\pd y}&=&
\left[-1+2\e-4\e\frac{y}{x+y+m^2}\right]\ov{A}_m
-2\frac{[m^2]^{-1-\e}}{(4\pi)^{2-\e}}
\left\{\frac{y}{x+y+m^2}X
{}_{2}F_1\left(1,1+\e;3/2-\e;-\frac{y}{m^2}\right)\right.\nonumber\\
&-&\left.\frac{Y}{2-\e} 
\left[{}_{2}F_1\left(2,1+\e;3-\e;-\frac{x+y}{m^2}\right)
+(1-\e) {}_{2}F_1\left(1,1+\e;3-\e;-\frac{x+y}{m^2}\right)
\right]\right\},
\label{eq:as4de0}
\eea
where
\be
X=\frac{\G(1/2-\e)\G(1+\e)}{\G(3/2-\e)},
\;\;\;\;Y=\frac{\G(1-\e)\G(1+\e)}{\G(2-\e)}.
\ee

%----------------------------------------------------------------------------
\subsection{Solving the differential equations}
  
Let us first consider the integral $A_m$.  By the same method as in 
\cite{Watson:2007mz}, we make the following ansatz:
\be
A_m(x,y)=F_{Am}(x,y)G_{Am}(x,y)
\ee
such that
\bea
2x\frac{\pd F_{Am}}{\pd x}&=&\left[1-(2+4\e)\frac{x}{x+y+m^2}\right]F_{Am},
\label{eq:agpdxh}\\
2y\frac{\pd F_{Am}}{\pd y}&=&
\left[-1+2\e-(2+4\e)\frac{y}{x+y+m^2}\right]F_{Am},\label{eq:agpdyh}\\
F_{Am}2x\frac{\pd G_{Am}}{\pd x}
&=&2\frac{[m^2]^{-1-\e}}{(4\pi)^{2-\e}}
\left\{
\frac{y+m^2}{x+y+m^2}X
{}_{2}F_1\left(1,1+\e;3/2-\e;-\frac{y}{m^2}\right)\right.\nonumber\\
&-&
\left.\frac{\G(-\e)\G(1+\e)}{\G(2-\e)}
{}_{2}F_1\left(2,1+\e;2-\e;-\frac{x+y}{m^2}\right)
+Y{}_{2}F_1\left(1,1+\e;2-\e;-\frac{x+y}{m^2}\right)\right\},
\label{eq:agpdx}\\
F_{Am}2y\frac{\pd G_{Am}}{\pd y}
&=&2\frac{[m^2]^{-1-\e}}{(4\pi)^{2-\e}}
\left\{-\frac{y}{x+y+m^2}X
{}_{2}F_1\left(1,1+\e;3/2-\e;-\frac{y}{m^2}\right)\right.\nonumber\\
&+&
\left.\frac{\G(-\e)\G(1+\e)}{\G(2-\e)}
{}_{2}F_1\left(2,1+\e;2-\e;-\frac{x+y}{m^2}\right)
+Y {}_{2}F_1\left(1,1+\e;2-\e;-\frac{x+y}{m^2}\right)\right\}.
\label{eq:agpdy}
\eea
By inspection, it is simple to determine the solution for the two 
homogeneous equations, \eq{eq:agpdxh} and \eq{eq:agpdyh}:
\be
F_{Am}(x,y)=x^{1/2}y^{-1/2+\e}(x+y+m^2)^{-1-2\e}.
\label{eq:af}
\ee
Since the mass $m$ is treated as a parameter, the (dimensionful) 
solution, \eq{eq:af}, may have an integration constant proportional 
to $[m^2]^{-1-\e}$. However, returning to the original equations, 
\eq{eq:agpdxh} and \eq{eq:agpdyh}, we see that the only consistent solution 
is the one for which this constant vanishes.

For the function $G_{Am}$ we make the following ansatz, which will be 
verified below ($z=x/y$): 
\be
G_{Am}(x,y)=G^0_{Am}(x,y)+\tilde{G}_{Am}(z).
\label{eq:ag}
\ee
The component $G^0_{Am}(x,y)$ can be found by adding the differential 
equations \eq{eq:agpdx} and \eq{eq:agpdy}, which lead to:
\bea
x\frac{\pd G^0_{Am}}{\pd x}+y\frac{\pd G^0_{Am}}{\pd y}
&=&\frac {1}{(4\pi)^{2-\e}}\left\{\frac{m^2}{\sqrt{x(y+m^2)}}
\ln\left(\frac{\sqrt{1+\frac{m^2}{y}}+1}{\sqrt{1+\frac{m^2}{y}}-1}\right)
+{\cal O}(\e)\right\}.
\eea
Because the function $G^0_{Am}$ is multiplied by the function $F_{Am}$ 
(which does not have an $\e$ pole), the term of order ${\cal O}(\e)$ will 
not contribute.  The solution of this equation is:
\be
G^0_{Am}(x,y)=-\frac{2}{(4\pi)^{2-\e}}\left\{\frac{1}{\sqrt{z}}
\left[\sqrt{1+\frac{m^2}{y}}\ln\left(\frac{\sqrt{1+\frac{m^2}{y}}+1}
{\sqrt{1+\frac{m^2}{y}}-1}\right)-\ln y\right]
+{\cal O}(\e)\right\}+{\cal C}_1.
\label{eq:agcomp0}
\ee

Before we proceed to determine $\tilde G_{Am}$, we justify the ansatz for 
$G_{Am}$, given by \eq{eq:ag}. First we observe that:
\be
2x F_{Am}\frac{\pd G_{Am}}{\pd x}=2zF_{Am}\frac{\pd \tilde G_{Am}}{\pd
  z}+ 2xF_{Am}\frac{\pd G^0_{Am}}{\pd x}.
\label{eq:agcomp0til}
\ee
Then we subtract the above equation, \eq{eq:agcomp0til}, from 
\eq{eq:agpdx}.  This gives:
\bea
 z\frac{\pd \tilde G_{Am}}{\pd z}&=&
\frac{1}{F_{Am}}\frac{[m^2]^{-1-\e}}{(4\pi)^{2-\e}}
\left\{
\frac{y+m^2}{x+y+m^2}X{}_{2}F_1\left(1,1+\e;3/2-\e;-\frac{y}{m^2}\right)
+Y{}_{2}F_1\left(1,1+\e;2-\e;-\frac{x+y}{m^2}\right)\right.
\nonumber\\
&-&\left.
\frac{\G(-\e)\G(1+\e)}{\G(2-\e)}
{}_{2}F_1\left(2,1+\e;2-\e;-\frac{x+y}{m^2}\right)
\right\}
-x \frac{\pd G^0_{Am}}{\pd x}.
\label{eq:tildegz}
\eea
Evaluation to the first order in $\e$ is straightforward and we see that 
the right hand side of the above expression is only a function of the 
variable $z$.  This allows us to write down a differential equation for 
$\tilde G_{Am} (z)$ in the form:
\bea
z\frac{\pd \tilde G_{Am}}{\pd z}
&=&\frac{1}{(4\pi)^{2-\e}}\frac{1}{\sqrt{z}}
\left\{\frac 1\e-\gamma+\ln m^2+\cal O(\e)\right\},
\eea
from which we get immediately
\be
\tilde G_{Am}(z)=-\frac{2}{(4\pi)^{2-\e}}
\frac{1}{\sqrt{z}}\left\{\frac 1\e -\gamma+\ln m^2+\cal O(\e)\right\}
+{\cal C}_2.
\label{eq:agtilde}
\ee
Returning to the original differential equations 
(\ref{eq:agpdxh} - \ref{eq:agpdy}) with the function 
$G(x,y)= G^0_{Am}(x,y)+\tilde G_{Am}(z)$, we see that the only consistent 
solution is the one for which the overall constant ${\cal C}_1+{\cal C}_2$ 
vanishes.

We may now put together the solutions Eqs.~(\ref{eq:af}), (\ref{eq:agcomp0}) 
and (\ref{eq:agtilde}) and write for the function $A_{m}$:
\bea
A_m(x,y)
&=&\frac{(x+y+m^2)^{-1-\e}}{(4\pi)^{2-\e}}
\left\{-\frac {2}{\e} +2\gamma+2\ln \left(\frac{x+y+m^2}{m^2}\right)
-2\sqrt{1+\frac{m^2}{y}}
\ln\left(\frac{\sqrt{1+\frac{m^2}{y}}+1}{\sqrt{1+\frac{m^2}{y}}-1}\right)
+\cal O(\e)\right\}.\nonumber\\
\label{eq:asol}
\eea
We see that for $m^2=0$ we regain the result from \cite{Watson:2007mz} and 
that the singularities are located at $x+y+m^2=0$ (with $m^2, y \ge 0$).  
Two more useful checks arise from the study of the power expansion around 
$x=0$ and the mass differential equation (for details, see 
Appendix~\ref{app:intcheck}).

We now proceed in the same fashion to determine the function 
$\ov{A}_m(x,y)=F_{\ov{A}m}(x,y)G_{\ov{A}m}(x,y)$.  The resulting partial 
differential equations are in this case:
\bea
2x\frac{\pd F_{\ov{A}m}}{\pd x}&=&
\left[-1-4\e\frac{x}{x+y+m^2}\right]F_{\ov{A}m},\\
2y\frac{\pd F_{\ov{A}m}}{\pd y}&=&
\left[-1+2\e-4\e\frac{y}{x+y+m^2}\right]F_{\ov{A}m},\\
F_{\ov{A}m}2x\frac{\pd G_{\ov{A}m}}{\pd x}&=&
2\frac{[m^2]^{-1-\e}}{(4\pi)^{2-\e}}
\left\{\frac{y+m^2}{x+y+m^2}X
{}_{2}F_1\left(1,1+\e;3/2-\e;-\frac{y}{m^2}\right)\right.\nonumber\\
&-&\left.\frac{Y}{2-\e}
\left[{}_{2}F_1\left(2,1+\e;3-\e;-\frac{x+y}{m^2}\right)
+(1-\e){}_{2}F_1\left(1,1+\e;3-\e;-\frac{x+y}{m^2}\right)
\right]\right\},
\label{eq:a4gpdx}\\
F_{\ov{A}m}2y\frac{\pd G_{\ov{A}m}}{\pd y}&=&
2\frac{[m^2]^{-1-\e}}{(4\pi)^{2-\e}}\left\{
-\frac{y}{x+y+m^2}X
{}_{2}F_1\left(1,1+\e;3/2-\e;-\frac{y}{m^2}\right)\right.\nonumber\\
&+&\left.\frac{Y}{2-\e} 
\left[{}_{2}F_1\left(2,1+\e;3-\e;-\frac{x+y}{m^2}\right)
+(1-\e){}_{2}F_1\left(1,1+\e;3-\e;-\frac{x+y}{m^2}\right)
\right]\right\},
\label{eq:a4gpdy}
\eea
with $X,Y$ defined previously.  The solution to the first pair is
\be
F_{\ov{A}m}(x,y)=x^{-1/2}y^{-1/2+\e}(x+y+m^2)^{-2\e}.
\label{eq:a4fsol}
\ee
For brevity, in the above expression and also in the derivation of the 
function $G_{\ov{A}m}=G^0_{\ov{A}m}+\tilde{G}_{\ov{A}m}$ (the analogue of 
$G_{Am}$) we omit the constants of integration -- they vanish as in the 
case of the functions $F_{Am}$ and $G_{Am}$. 

As before, for $G_{\ov{A}m}(x,y)$ we make the ansatz:
\be
G_{\ov{A}m}(x,y)=G^0_{\ov{A}m}(x,y)+\tilde{G}_{\ov{A}m}(z).
\label{eq:a4g}
\ee
In the limit $\e\rightarrow 0$, the component $G^0_{\ov{A}m}(x,y)$ is 
determined from the differential equation:
\bea
x\frac{\pd G^0_{\ov{A}m}}{\pd x}+y\frac{\pd G^0_{\ov{A}m}}{\pd y}&=&
\frac{1}{(4\pi)^{2-\e}}
\left\{\frac{\sqrt{x}}{x+y+m^2}\frac{m^2}{\sqrt{y+m^2}} 
\ln\left(\frac{\sqrt{1+\frac{m^2}{y}}+1}{\sqrt{1+\frac{m^2}{y}}-1}\right)
+{\cal O}(\e)\right\}.
\eea
The solution of this equation is:
\bea
&&G^0_{\ov{A}m}(x,y)=\frac{1}{(4\pi)^{2-\e}}
\left\{\imath\ln\left(\frac{\sqrt{1+\frac{m^2}{y}}-
\imath\sqrt{z}}{\sqrt{1+\frac{m^2}{y}}+\imath\sqrt{z}}\right)
 \ln\left(\frac{\imath\sqrt{z}+1}{\imath\sqrt{z}-1}\right)\right.\nonumber\\
&&\left.-\imath \textrm{Li}_2\left(\frac{1-\imath\sqrt{z}}{1+\imath\sqrt{z}}
\cdot
\frac{\sqrt{1+\frac{m^2}{y}}-\imath\sqrt{z}}{\sqrt{1+\frac{m^2}{y}}
+\imath\sqrt{z}}
\right)
+\imath \textrm{Li}_2 \left(\frac{1+\imath\sqrt{z}}{1-\imath\sqrt{z}}\cdot
\frac{\sqrt{1+\frac{m^2}{y}}-\imath\sqrt{z}}{\sqrt{1+\frac{m^2}{y}}
+\imath\sqrt{z}}
\right)+{\cal O}(\e)\right\},
\label{eq:a4gcomp0}
\eea
where $\textrm{Li}_2(z)$ is the dilogarithmic function \cite{lewin}:
\be
\textrm{Li}_2(z)=-\int_0^z\frac{\ln(1-t)}{t}dt.
\ee

As before, we check that the ansatz for $G_{\ov{A}m}(x,y)$ given in 
\eq{eq:a4g} is correct and derive the differential equation for the 
function $\tilde G_{\ov{A}m}$, in the limit $\e\rightarrow 0$:
\be
z \frac{\pd\tilde G_{\ov{A}m}}{\pd
  z}=\frac{1}{(4\pi)^{2-\e}}\left\{\frac{\sqrt{z}}{z+1}\left[ \ln
    (1+z)-\ln z-2\ln 2\right]+{\cal O}(\e)\right\}.
\ee
The result we leave for the moment in the form:
\bea
\tilde G_{\ov{A}m}(z)=\frac{1}{(4\pi)^{2-\e}}
\left\{-4\ln 2\arctan(\sqrt{z})+\int_0^z\frac{dt}{\sqrt{t}(1+t)}\ln{(1+t)}
-\int_0^z\frac{dt}{\sqrt{t}(1+t)}\ln{t}+{\cal O}(\e)\right\}.
\label{eq:a4tildegzsol}
\eea
With the solutions, Eqs.~(\ref{eq:a4fsol}), (\ref{eq:a4gcomp0}) and 
(\ref{eq:a4tildegzsol}), after some further manipulation we can write down 
the following simplified expression for the integral $A_m^4$:
\bea
A^4_m(x,y)&=&k_4\frac{(x+y+m^2)^{-1-\e}}{(4\pi)^{2-\e}}
\frac{(1+z+\frac{m^2}{y})}{\sqrt{z}}
\left\{
2\ln\left(\!\!\frac{m^2}{y}\!\right)\arctan{(\sqrt{z})}
+2\ln\left(\!\frac{\sqrt{1+\frac{m^2}{y}}+1}{\sqrt{1+\frac{m^2}{y}}-1}\right)
\arctan{\left(\frac{\sqrt{z}}{\sqrt{\frac{m^2}{y}+1}}\right)}
\right.\nonumber\\
&-&\left.\int_0^z\frac{dt}{\sqrt{t}(1+t)}\ln{\left(1+t+\frac{m^2}{y}\right)}
+{\cal O}(\e)\right\},
\label{eq:a4sol}
\eea
with  the integral
\bea
&&\int_0^z\frac{dt}{\sqrt{t}(1+t)}\ln{\left(1+t+\frac{m^2}{y}\right)}=
\pi\ln{2}
-\imath \ln\left(\frac{1-\imath\sqrt{z}}{1+\imath\sqrt{z}}\right)2\ln 2
\nonumber\\
&+&\imath \ln\left(\frac{1-\imath\sqrt{z}}{1+\imath\sqrt{z}}
\frac{\sqrt{1+\frac{m^2}{y}}-\imath\sqrt{z}}{\sqrt{1+\frac{m^2}{y}}
+\imath\sqrt{z}}\right)
\ln\left(\sqrt{1+\frac{m^2}{y}}+1\right)
+\imath \ln\left(\frac{1-\imath\sqrt{z}}{1+\imath\sqrt{z}}
\frac{\sqrt{1+\frac{m^2}{y}}+\imath\sqrt{z}}{\sqrt{1+\frac{m^2}{y}}
-\imath\sqrt{z}}\right)
\ln\left(\sqrt{1+\frac{m^2}{y}}-1\right)\nonumber\\
&-&\imath\ln{(\sqrt{z}-\imath)}\left[\ln{2}+\ln{(1+z+\frac{m^2}{y})}
-\ln{(1-\imath\sqrt{z})}-\frac12\ln{(\sqrt{z}-\imath)}
\right]
-\imath \textrm{Li}_2{\left(\ha-\frac{\imath}{2}\sqrt{z}\right)}
+\imath \textrm{Li}_2{\left(\ha+\frac{\imath}{2}\sqrt{z}\right)}
\nonumber\\
&+&\imath\ln{(\sqrt{z}+\imath)}\left[\ln{2}+\ln{(1+z+\frac{m^2}{y})}
-\ln{(1+\imath\sqrt{z})}-\ha\ln{(\sqrt{z}+\imath)}
\right]
-\imath\textrm{Li}_2\left(\frac{\imath+\sqrt{z}}{-\imath+\sqrt{z}}\right)
+\imath\textrm{Li}_2\left(\frac{-\imath+\sqrt{z}}{\imath+\sqrt{z}}\right)
\nonumber\\
&+&\imath\textrm{Li}_2
\left(\frac{\imath\sqrt{1+\frac{m^2}{y}}+\sqrt{z}}{-\imath+\sqrt{z}}\right)
-\imath\textrm{Li}_2
\left(\frac{-\imath\sqrt{1+\frac{m^2}{y}}+\sqrt{z}}{\imath+\sqrt{z}}\right)
+\imath\textrm{Li}_2
\left(\frac{-\imath\sqrt{1+\frac{m^2}{y}}+\sqrt{z}}{-\imath+\sqrt{z}}\right)
+\imath\textrm{Li}_2
\left(\frac{\imath\sqrt{1+\frac{m^2}{y}}+\sqrt{z}}{\imath+\sqrt{z}}\right).
\nonumber\\
\eea

We see that for $m^2=0$ we get the correct limit for the function $A_m^4$.  
We also mention that the singularities are located at $x+y+m^2=0$ and the 
apparent singularities at $z=-1$ (i.e., $x+y=0)$ in the expression 
\eq{eq:a4sol} are canceling out.  This can be easily seen by making a 
series expansion of \eq {eq:a4sol} around $z=-1$:
\be
A^4_m\stackrel{z\rightarrow-1}{=}\frac{1}{\sqrt{z}}
\left[\frac{y}{m^2}-\frac{z+1}{2}\left(\frac{y}{m^2}\right)^2
+{\cal O}\left((z+1)^3\right)
\right]
-\frac{\sqrt{1+\frac{m^2}{y}}}{\sqrt{z}(z+1+\frac{m^2}{y})}
\ln\left(\frac{\sqrt{1+\frac{m^2}{y}}+1}{\sqrt{1+\frac{m^2}{y}}-1}\right)
+\cal O(\e).
\label{eq:a4expz-1}
\ee
Again, the result \eq{eq:a4sol} has been checked by performing an expansion 
around $x=0$ and by studying the mass differential equation (see 
Appendix~\ref{app:intcheck}).

%-----------------------------------------------------------------------------
\section{Perturbative results in the limit $\e\rightarrow 0$}
\setcounter{equation}{0}

We can now collect together the results and write down the one-loop 
perturbative expressions for the two-point functions.  In 
Eqs.~(\ref{eq:atmassive}), (\ref{eq:bmassive}) and (\ref{eq:asmassive1}) 
we insert the corresponding integrals (derived in the previous section and 
Appendix~\ref{app:stdint}) and find for the temporal, spatial and massive 
components of the quark gap equation, in the limit $\e\rightarrow 0$:
\bea
\lefteqn{
A_t(k)=1}\nonumber\\&&
+\frac{C_F g^2}{(4\pi)^{2-\e}}\left\{
\frac{1}{\e}-\ga-\ln\frac{m^2}{\mu}+1-\frac{m^2}{k^2}
+\left(\frac{m^4}{k^4}-1\right)\ln\left( 1+\frac{k^2}{m^2}\right)
+\cal{O}(\e)
\right\},
\label{eq:atmassive_eps}\\
\lefteqn{A_s(k)=1}\nonumber\\&&
+\frac{C_Fg^2}{(4\pi)^{2-\e}}
\left\{
\frac1\e-\ga-\ln\frac{m^2}{\mu}+1+8\frac{k^2}{\vec k^2}
+4\frac{m^2}{\vec k^2}-\frac{m^2}{k^2}\right.
+\left(1+\frac{m^2}{k^2}\right)\left(4\frac{k^2}{\vec k^2}-1
+\frac{m^2}{k^2}\right)
\ln \left( 1+\frac{k^2}{m^2}\right)\nonumber\\
&&-\left.\left(4\frac{k^2}{\vec k^2}+2\frac{m^2}{\vec k^2}\right)
\sqrt{1+\frac{m^2}{\vec k^2}}
\ln\left(\frac{\sqrt{1+\frac{m^2}{\vec k^2}}+1}
{\sqrt{1+\frac{m^2}{\vec k^2}}-1}\right)
-2\frac{k_4^2}{\vec k^4}(k^2+m^2)f_m\left(k_4^2,\vec{k}^2\right)
+\cal{O}(\e)\right\},
\label{eq:asmassive_eps}\\
\lefteqn{B_m(k)=m}\nonumber\\&&
+m\frac{C_Fg^2}{(4\pi)^{2-\e}}\left\{
\frac4\e-4\ga-4\ln\frac{m^2}{\mu}+10-2\sqrt{1+\frac{m^2}{\vec k^2}}
\ln\left(\frac{\sqrt{1+\frac{m^2}{\vec k^2}}+1}
{\sqrt{1+\frac{ m^2}{\vec k^2}}-1}\right)
-2\left(1+\frac{m^2}{k^2}\right)\ln\left(1+\frac{k^2}{m^2}\right)+\cal{O}(\e)
\right\},\nonumber\\
\label{eq:bmassive_eps}
\eea
where the function $f_m(x,y)$ is given by ($z=x/y\equiv k_4^2/\vec k^2$):
\bea
f_m(x,y)&=&\frac{2}{\sqrt z}\ln\left(\frac{m^2}{y}\right)\arctan{(\sqrt{z})}
+\frac{2}{\sqrt z}
\ln\left(\frac{\sqrt{1+\frac{m^2}{y}}+1}{\sqrt{1+\frac{m^2}{y}}-1}\right)
\arctan{\left(\frac{\sqrt{z}}{\sqrt{\frac{m^2}{y}+1}}\right)}\nonumber\\
&-&\int_0^1\frac{dt}{\sqrt{t}(1+zt)}\ln{\left(1+zt+\frac{m^2}{y}\right)}.
\eea
The last integral has been rewritten using the identity:
\be
\frac{1}{\sqrt{z}}\int_0^z\frac{dt}{\sqrt{t}(1+t)}
\ln{\left(1+t+\frac{m^2}{y}\right)}=
\int_0^1\frac{dt}{\sqrt{t}(1+zt)}\ln{\left(1+zt+\frac{m^2}{y}\right)}.
\ee
As a useful check, we can set $m=0$ and show that the results for the 
temporal and spatial components are in agreement with the calculation 
performed independently using the one-loop massless integrals derived in 
Ref.~\cite{Watson:2007mz}.  As has been shown in the previous section, in the 
noncovariant integrals the singularities appear at $x+y+m^2=0$. It is easy 
to see that the standard integrals also have the same singularity 
structure.  Because of the absence of singularities in both the Euclidean 
and spacelike Minkowski regions, we can see that the validity of the Wick 
rotation is justified.

Having calculated the dressing functions for the quark proper two-point 
Green's function, we are now able to discuss the structure of the 
propagator.  In \eq{eq:b} we first analyze the denominator factor.  Let us 
denote (in Euclidean space):
\be
D(k)=k_4^2A_t^2(k)+\vec k^2A_s^2(k)+B_m^2(k).
\ee
Inserting the expressions from Eqs.~(\ref{eq:atmassive_eps}), 
(\ref{eq:asmassive_eps}) and (\ref{eq:bmassive_eps}) into the above 
equation, we have:
\bea
\lefteqn{D(k)
=k^2+m^2\left\{1+6\frac{g^2C_F}{(4\pi)^{2-\e}}
\left[\frac1\e-\ga-\ln\frac{m^2}{\mu}+\frac43\right]\right\}
+(k^2+m^2)\frac{2C_F g^2}{(4\pi)^{2-\e}}
\left\{\frac1\e-\ga-\ln\frac{m^2}{\mu}\right\}}\nonumber\\&&
+(k^2+m^2)\frac{2C_F g^2}{(4\pi)^{2-\e}}
\left\{9+\left(3-\frac{m^2}{k^2}\right)\ln\left(1+\frac{k^2}{m^2}\right)
-4\sqrt{1+\frac{m^2}{\vec k^2}}
\ln\left(\frac{\sqrt{1+\frac{m^2}{\vec k^2}}+1}{\sqrt{1+\frac{m^2}{\vec
        k^2}}-1}\right)-2\frac{k_4^2}{\vec k^2}f_m(k_4^2,\vec k^2) \right\}.
\label{eq:denfact}
\eea
We define the renormalized mass, $m_R$, via:
\be
m^2=Z_m^2 m_R^2
\textrm{~~~with~~~} 
Z_m^2=1-6\frac{g^2C_F}{(4\pi)^{2-\e}}
\left\{\frac1\e-\ga-\ln\frac{m^2}{\mu}+\frac43\right\}.
\ee
The expression for $D(k)$, \eq{eq:denfact}, then contains explicitly the 
overall factor $k^2+m_R^2$, meaning that the simple pole mass of the quark 
emerges, just as it does in covariant gauges. The singularity structure of 
the remaining part is such that there are no non-analytic structures for 
spacelike or Euclidean momenta.  Moreover, we see that the renormalization 
factor, $Z_m$, which should be a gauge invariant quantity (it defines the 
physical perturbative pole mass) agrees with the result obtained in 
covariant gauges \cite{muta}.

Because of the Dirac structure, it is more convenient to write the quark 
propagator in Minkowski space. We have shown that the analytic continuation 
of the functions $A_t,A_s, B_m$ back into the Minkowski space is allowed 
and this enables us simply to write (note that also in $D(k)$ we must also 
analytically continue $k_4^2\rightarrow-k_0^2$):
\be
W_{\bar qq}(k)
=\imath\left\{\ga^0k_0A_t(k)-\ga^ik_iA_s(k)+B_m(k)\right\} D^{-1}(k). 
\ee
Inserting the denominator factor, \eq{eq:denfact}, in the limit 
$\e\rightarrow 0$ and replacing the mass with its renormalized counterpart, 
the above expression gives: 
\be
W_{\bar qq}(k)=-\frac{i}{k_0^2-\vec k^2-m_R^2}\left\{(\kslash+m_R)
\left[1-C_F\frac{g^2}{(4\pi)^{2-\e}}\left(\frac1\e-\ga\right)\right]
+\textrm{~finite~ terms~} \right\}.
\ee
We can thus write down for the quark propagator:
\be
W_{\bar qq}(k)=-\frac{\imath(\kslash+m_R)} {k_0^2-\vec k^2-m_R^2}Z_2
+\textrm{~finite~ terms~}
\ee
and identify the renormalization constant (omitting the prescription 
dependent constants)
\be
Z_2=1-\frac{g^2C_F}{(4\pi)^{2-\e}}\left(\frac1\e-\ga\right).
\ee

Turning to the quark loop contributions to the gluon two-point proper 
functions, in evaluating the integral structure of Eqs.~(\ref{eq:Gsisi}), 
(\ref{eq:GsiA}) and (\ref{eq:GAA}) we observe the following relations (in 
Euclidean space): 
\bea
\G_{\si\si(q)}^{(1)}(k)=\G_{\si A(q)}^{(1)}(k)=
-\frac{\vec k^2}{k^2}\G_{AA(q)}^{(1)}(k)
=-\frac{\vec k^2}{k_4^2}\bar\G_{AA,q}^{(1)}(k)=I(k_4^2,\vec k^2),
\eea
where the integral $I(k_4^2,\vec k^2)$ reads (using the results of 
Appendix~\ref{app:stdint}), as $\e\rightarrow 0$:
\bea
I(k_4^2,\vec k^2)&=&\frac{N_f}{(4\pi)^{2-\e}}
\left\{
-\frac23\left[\frac{1}{\e}-\ga-\ln \frac{k^2}{\mu}\right]-\frac{10}{9}
+\frac23\sqrt{1+\frac{4m^2}{k^2}}\left(1-2\frac{m^2}{k^2}\right)
\ln
\left(\frac{\sqrt{1+\frac{4m^2}{k^2}}+1}{\sqrt{1+\frac{4m^2}{k^2}}-1}\right)
\right.\nonumber\\
&+&\left.\frac23\left(4\frac{m^2}{k^2}+\ln \frac{m^2}{k^2}\right)
+\cal O(\e)\right\}.
\eea
The above integral agrees with the results obtained in covariant gauges 
(see for instance \cite{muta}).  This is hardly surprising, since at 
one-loop level the quark loop as a whole is unchanged from its covariant 
counterpart --- what is different is that the various degrees of freedom 
(temporal and spatial) are being separated into the corresponding proper 
two-point functions, i.e., $\G_{AA}$, $\G_{A\si}$ and $\G_{\si\si}$.

The one-loop gluon propagator dressing functions we construct by writing 
$D=D^{(0)}+g^2D^{(1)}$.  As mentioned previously, in the first order 
formalism we have to account for the presence of the additional 
$\vec\pi, \phi$ and ghost fields and the corresponding propagators 
(for example $D_{A\pi}$). Whilst the quarks only contribute to three of the 
gluon proper two-point functions ($\G_{AA},\G_{A\si}$ and $\G_{\si\si}$) at 
one-loop, there will be contributions to many more of the various connected 
(propagator) two-point functions. The relationship between the connected and 
proper gluon two-point functions in the first order formalism is detailed in 
Ref.~\cite{Watson:2007mz}. The full set of quark contributions to these 
gluonic type propagators is: 
\bea
&&D_{AA(q)}^{(1)}(k)=D_{\si\si(q)}^{(1)}(k)=\G_{\si\si(q)}^{(1)}(k)=
I(k_4^2,\vec k^2),\nonumber\\
&&D_{A\pi(q)}^{(1)}(k)=-\frac{\vec k^2}{k_4^2}D_{\pi\pi(q)}^{(1)}(k)
=D_{\si\phi(q)}^{(1)}(k)=-D_{\phi\phi(q)}^{(1)}(k)= I(k_4^2,\vec k^2).
\label{eq:letransf}
\eea

At this point we are able to identify the first coefficient of the 
perturbative $\ba$-function.  As is well known in Landau gauge, a 
renormalization group invariant running coupling can be defined through 
the following perturbative combination of gluon and ghost propagator 
dressing functions \cite{Fischer:2006ub}:
\be
g^2D_{AA}D_c^{2}\sim g^2\left[1+\frac{g^2}{16\pi^2}\frac{1}{\e}
\left(\frac{11N_c}{3}-\frac{2N_f}{3}\right)\right].
\ee
At one-loop in perturbation theory, the coefficient of the $1/\e$ pole 
above is simply minus the first coefficient of the $\ba$-function 
($\beta_0=-11N_c/3+2N_f/3$).  By inspecting the relations \eq{eq:letransf} 
and those obtained in Ref.~\cite{Watson:2007mz} for the propagators 
$D_{AA}$ and $D_{c}$, we see that the same result is achieved in Coulomb 
gauge.  In Coulomb gauge, a second renormalization group invariant 
combination of propagators appears and is given by $g^2 D_{\si\si}$ 
\cite{Zwanziger:1998ez}.  Again, combining our results \eq{eq:letransf} 
and those obtained in \cite{Watson:2007mz}, we see that indeed the 
coefficient of $1/\e$ agrees with this.

%-----------------------------------------------------------------------------
\section{Summary and Outlook}
\setcounter{equation}{0}

In this paper, the quark contributions to the \DS equations of QCD have 
been derived within the Coulomb gauge first order formalism and perturbative 
results have been presented.  The set of Feynman rules has been derived and 
the general form of the two-point functions have been established.  The 
quark gap equation and the quark loop contributions to the \DS equations 
concerning the gluon proper two-point functions have been explicitly 
derived.  A one-loop perturbative calculation has been performed, for the 
quark gap equation, as well as for the quark contributions to the gluon 
proper two-point functions.  The required noncovariant massive integrals 
have been obtained, using techniques based on differential equations and 
integration by parts.  The various two-point dressing functions and 
propagators have been evaluated in the limit $\e\rightarrow 0$.  The 
validity of the analytic continuation between Minkowski and Euclidean 
space has been verified.   The quark mass and propagator have been 
renormalized and it is seen that the one-loop result for the gauge 
invariant quark mass renormalization coefficient agrees explicitly with 
the result obtained in linear covariant gauges.  The correct one-loop 
coefficient for the $\ba$ function has also been obtained.

The natural continuation of this work is the perturbative evaluation of 
vertex functions of the theory.  The Mellin-Barnes parametrization or 
perhaps a generalization of the differential equation method to the 
three-point integrals would be possible ways to proceed. Also, the 
construction of scattering matrix elements would be another interesting topic. 
 
%-----------------------------------------------------------------------------
\begin{acknowledgments}
The authors are grateful to Davide Campagnari for a critical reading of
the manuscript.
CP has been supported by the Deutscher Akademischer Austausch Dienst (DAAD).
PW and HR have been supported by the Deutsche Forschungsgemeinschaft
(DFG) under Contracts No. DFG-Re856/6-1 and DFG-Re856/6-2.
\end{acknowledgments}
%----------------------------------------------------------------------------

%----------------------------------------------------------------------------
\appendix
%----------------------------------------------------------------------------
%----------------------------------------------------------------------------
\section{Standard massive integrals }
\setcounter{equation}{0}
\label{app:stdint}

Consider the integral:
\be
J_m(k^2)=\int\frac{\dk{\w}}{[\w^2+m^2]^\mu[(k-\w)^2+m^2]^\nu}.
\label{eq:smi_example}
\ee
In the case $\mu=\nu=1$ this gives the scalar integral associated with, for 
example, the fermion loop in quantum electrodynamics \cite{muta}.  We 
present here a method to evaluate such integrals for arbitrary denominator 
powers (developed originally in Ref.~\cite{Boos:1990rg}) and generalize to 
the various additional noncovariant integrals.

We start by writing the Taylor expansion of the massive propagator in terms 
of a hypergeometric function in the following way:
\be
\frac{1}{[\w^2+m^2]^{\mu}}=\frac{1}{[\w^2]^{\mu}}
{}_{1}F_{0} \left(\mu;-\frac{m^2}{\w^2}\right).
\label{eq:prop}
\ee
Now, the idea is to use the Mellin-Barnes representation of the 
hypergeometric function ${}_{1}F_{0} (\mu;z)$:
\be
{}_{1}F_{0} (\mu;z)=
\frac{1}{\G(\mu)}\frac{1}{2\pi\imath}
\int\limits_{-\imath\infty}^{\imath\infty}d s (-z)^{s}\G(-s)\G(\mu+s),
\label{eq:hyprep}
\ee
where the contour in the complex plane separates the left poles of the 
$\G$ functions from the right poles. A first advantage of this 
representation is that the ``mass term'' gets separated from the massless 
propagator and the remaining integrals can be calculated with the Cauchy 
residue theorem, as we shall see below.  Also, we note that the results can 
be written as a function of  either $k^2/m^2$, or $m^2/k^2$ (expansions 
thereof are of interest in studying various momentum regimes). This we do 
by using the well-known formulas of analytic continuation of the 
hypergeometric function (for an extended discussion, see \cite{Boos:1990rg}). 

Applying \eq{eq:hyprep} to the massive propagator we can rewrite the 
integral $J_m(k^2)$ as: 
\bea
J_m(k^2)=
\frac{1}{(2\pi\imath)^2}\frac{1}{\G(\mu)\G(\nu)}
\int\!\!\!\!\!\int\limits_{-\imath\infty}^{\imath\infty}ds dt 
(m^2)^{s+t}\G(-s)\G(-t)\G(\mu+s)\G(\nu+t)
\int\frac{\dk{\w}}{(\w^2)^{\mu+s}[(k-\w)^2]^{\nu+t}}.
\eea
Inserting the general result for the massless integral (an explicit 
derivation can be found in Ref.~\cite{Watson:2007mz}):
\be
\int\frac{\dk{\w}}{[\w^2]^\mu[(k-\w)^2]^\nu}
=\frac{[k^2]^{2-\mu-\nu-\e}}{(4\pi)^{2-\e}}
\frac{\G(\mu+\nu+\e-2)}{\G(\mu)\G(\nu)}
\frac{\G(2-\mu-\e)\G(2-\nu-\e)}{\G(4-\mu-\nu-2\e)},
\label{eq:si_ex}
\ee
we get for the integral $J_m$:
\bea
J_m(k^2)&=&\frac{[k^2]^{2-\nu-\mu-\e}}{(4\pi)^{2-\e}}
\frac{1}{(2\pi\imath)^2}\frac{1}{\G(\mu)\G(\nu)}
\int\!\!\!\!\!\int\limits_{-\imath\infty}^{\imath\infty}
\!\!ds dt \left(\frac{m^2}{k^2}\right)^{s+t}
\G(-s)\G(-t)\G(2-\e-\mu-s)\G(2-\e-\nu-t)\nonumber\\ 
&&\times\frac{\G(\mu+\nu+s+t-2+\e)}{\G(4-2\e-\mu-\nu-s-t)}.
\eea
With the change of variable $t=2-\e -\mu-\nu-u-s$ (for such a replacement, 
the left and right poles of the $\G$ function are simply interchanged and 
therefore the condition of separating the poles is not contradicted) we 
obtain:
\bea
J_m(k^2)&=&\frac{[m^2]^{2-\mu-\nu-\e}}{(4\pi)^{2-\e}}
\frac{1}{(2\pi\imath)}\frac{1}{\G(\mu)\G(\nu)}
\int\limits_{-\imath\infty}^{\imath\infty}du
\left(\frac{m^2}{k^2}\right)^{-u}\!\frac{\G(-u)}{\G(2-\e+u)}\nonumber\\
&&\times\frac{1}{(2\pi\imath)}\int\limits_{-\imath\infty}^{\imath\infty}ds 
\G(-s)\G(2-\e-\mu-s)\G(-2+\e+\nu+\mu+u+s)\G(\mu+u+s).
\label{eq:Jintermediate}
\eea
To evaluate the integral over $s$ we use the Barnes Lemma:
\be
\frac{1}{(2\pi\imath)}\int\limits_{-\imath\infty}^{\imath\infty}ds\textrm{~}
\G(a+s)\G(b+s)\G(c-s)\G(d-s)=
\frac{\G(a+c)\G(a+d)\G(b+c)\G(b+d)}{\G(a+b+c+d)}
\ee
and for the integral \eq{eq:Jintermediate} it follows immediately that
\bea
J_m(k^2)&=&\frac{[m^2]^{2-\mu-\nu-\e}}{(4\pi)^{2-\e}}
\frac{1}{(2\pi\imath)}\frac{1}{\G(\mu)\G(\nu)}
\int\limits_{-\imath\infty}^{\imath\infty}\textrm{d}u
\left(\frac{m^2}{k^2}\right)^{-u}
\frac{\G(-u)\G(\mu+u)\G(\nu+u)\G(\mu+\nu-2+\e+u)}{\G(\mu+\nu+2u)}. 
\eea
Closing the integration contour on the right we have:
\bea
J_m(k^2)&=&\frac{[m^2]^{2-\mu-\nu-\e}}{(4\pi)^{2-\e}}
\frac{1}{(2\pi\imath)}\frac{1}{\G(\mu)\G(\nu)}(2\pi\imath)
\sum\limits_{j=0}^{\infty}
\left(-\frac{m^2}{k^2}\right)^{-j}\frac{1}{j!}
\frac{\G(\mu+j)\G(\nu+j)\G(\mu+\nu-2+\e+j)}{\G(\mu+\nu+2j)} .
\eea
With the help of the duplication formula
\be
\G(2z)=2^{2z-1}\pi^{-1/2}\G(z)\G\left(z+\frac 12\right),
\ee
we can rewrite $J_m$ as:
\bea
J_m(k^2)&=&\frac{[m^2]^{2-\mu-\nu-\e}}{(4\pi)^{2-\e}}
\frac{\G(\mu+\nu-2+\e)}{\G(\mu+\nu)}\nonumber\\
&&\sum\limits_{j=0}^{\infty}
\left(-\frac{m^2}{k^2}\right)^{-j}\frac{1}{2^{2j}}\frac{1}{j!}
\frac{\G(\mu+j)}{\G(\mu)}\frac{\G(\nu+j)}{\G(\nu)}
\frac{\G(\mu+\nu-2+\e+j)}{\G(\mu+\nu-2+\e)}
\frac{\G(\frac{\mu+\nu}{2})} {\G(\frac{\mu+\nu}{2}+j)}
\frac{\G(\frac{\mu+\nu+1}{2})} {\G(\frac{\mu+\nu+1}{2}+j)}.
\eea
The sum is clearly a representation of the hypergeometric 
${}_{3}F_{2}(a,b,c;d,e;z)$ (see \cite{abramowitz}) and we finally obtain:
\bea
J_m(k^2)&=&\frac{[m^2]^{2-\mu-\nu-\e}}{(4\pi)^{2-\e}}
\frac{\G(\mu+\nu-2+\e)}{\G(\mu+\nu)}
{}_{3}F_{2} \left(\mu,\nu,\mu+\nu-2+\e;\frac{\mu+\nu}{2},\frac{\mu+\nu+1}{2}
;-\frac{k^2}{4m^2}\right).
\label{eq:smi1}
\eea
A trivial computation shows that the result \eq{eq:smi1} is consistent with 
the known results in the limit $m=0$.  All we have to do is to invert the 
argument of the hypergeometric according to the formula (see, for example, 
Ref.~\cite{bateman})
\bea
&&{}_{3}F_{2}(a_1,a_2,a_3;b_1,b_2;z)=
\frac{\G(b_1)\G(b_2)}{\G(a_1)\G(a_2)\G(a_3)}
\left\{
\frac{\G(a_1)\G(a_2-a_1)\G(a_3-a_1)}{\G(b_1-a_1)\G(b_2-a_1)}(-z)^{-a_1}
\right.\nonumber\\
&&\times{}_{3}F_{2}
\left(a_1,a_1-b_1+1,a_1-b_2+1;a_1-a_2+1,a_1-a_3+1;\frac 1z\right)\nonumber\\
&+&\frac{\G(a_2)\G(a_1-a_2)\G(a_3-a_2)}{\G(b_1-a_2)\G(b_2-a_2)}(-z)^{-a_2}
{}_{3}F_{2}
\left(a_2,a_2-b_1+1,a_2-b_2+1;-a_1+a_2+1,a_2-a_3+1;\frac 1z\right)\nonumber\\
&+&\left.
\frac{\G(a_3)\G(a_1-a_3)\G(a_2-a_3)}{\G(b_1-a_3)\G(b_2-a_3)}(-z)^{-a_3}
{}_{3}F_{2}\left(a_3,a_3-b_1+1,a_3-b_2+1;-a_1+a_3+1,-a_2+a_3+1;\frac 1z\right)
\right\}.\nonumber\\
\eea

For integrals with different type of denominator factors, similar 
calculations bring us to the following results:
\bea
\int\frac{\dk{\w}}{[\w^2]^\mu[(k-\w)^2+m^2]^\nu}
\!&=\!&
\frac{[m^2]^{2-\mu-\nu-\e}}{(4\pi)^{2-\e}}
\frac{\G(2-\mu-\e) \G(\mu+\nu+\e-2)}{\G(\nu)\G(2-\e)}
{}_{2}F_{1} \left(\mu,\mu+\nu+\e-2;2-\e;-\frac{k^2}{m^2}\right),\nonumber\\
\label{eq:smi2}\\
\int\frac{\dk{\w}}{[\vec{\w}^2]^\mu[(k-\w)^2+m^2]^\nu}
\!&=\!&
\frac{[m^2]^{2-\mu-\nu-\e}}{(4\pi)^{2-\e}}
\frac{\G(\frac32-\mu-\e) \G(\mu+\nu+\e-2)}{\G(\nu)\G(3/2-\e)}
{}_{2}F_{1} \left(\mu,\mu+\nu+\e-2;3/2-\e;-\frac{\vec{k}^2}{m^2}\right).
\nonumber\\
\label{eq:smi3}
\eea
This method can also be applied to integrals with more complicated numerator 
structure.

For completeness, we also show the first order $\e$ expansion of the 
integrals arising into the one-loop perturbative expressions considered in 
this work:
\bea
\int\frac{\dk{\w}}{(\w^2+m^2)[(k-\w)^2+m^2]}
&=&\frac{[m^2]^{-\e}}{(4\pi)^{2-\e}}
\left\{
\frac1\e-\ga+2-\sqrt{1+\frac{4m^2}{k^2}}
\ln\left(\frac{\sqrt{1+\frac{4m^2}{k^2}}+1}{\sqrt{1+\frac{4m^2}{k^2}}-1}
\right)
+\cal{O}(\e)
\right\},\\
\int\frac{\dk{\w}}{\w^2[(k-\w)^2+m^2]}
&=&\frac{[m^2]^{-\e}}{(4\pi)^{2-\e}}
\left\{
\frac1\e-\ga+2-\left(1+\frac{m^2}{k^2}\right)\ln \left( 1+\frac{k^2}{m^2}
\right)+\cal{O}(\e)
\right\},\\
\int\frac{\dk{\w} \w_i}{\w^2[(k-\w)^2+m^2]}
&=&k_i\frac{[m^2]^{-\e}}{(4\pi)^{2-\e}}
\left\{
\frac12\left(\frac1\e-\ga\right)+1+\frac12\frac{m^2}{k^2}
-\frac12\left(1+\frac{m^2}{k^2}\right)^2\ln \left( 1+\frac{k^2}{m^2}
\right)+\cal{O}(\e)
\right\},\nonumber\\
\\
\int\frac{\dk{\w}}{\vec\w^2[(k-\w)^2+m^2]}
&=&\frac{[m^2]^{-\e}}{(4\pi)^{2-\e}}
\left\{
\frac2\e-2\ga+8-2\sqrt{1+\frac{m^2}{\vec k^2}}
\ln\left(\frac{\sqrt{1+\frac{m^2}{\vec k^2}}+1}
{\sqrt{1+\frac{m^2}{\vec k^2}}-1}\right)
+\cal{O}(\e)
\right\}.
\eea

%----------------------------------------------------------------------------
\section{Checking the Nonstandard Integrals}
\label{app:intcheck}
\setcounter{equation}{0}

One way to check analytically the results for the integrals $A_m$ and 
$A_m^4$, \eq{eq:asol} and \eq{eq:a4sol}, respectively, is to make an 
expansion around $x=0$ and evaluate the resulting integrals with the help 
of the Schwinger parametrization.  Let us consider then the integral $A_m$, 
originally defined in \eq{eq:adef}.  Using Schwinger parameters 
\cite{pascual}, we can rewrite the denominator factors as exponential 
functions to give:
\be
A_m=\int_0^\infty\,d\al d\ba d\ga \int\dk{\w}\exp{\left\{-(\al+\ba)\w_4^2
+2\ba k_4\w_4-\ba k_4^2-(\al+\ba+\ga)\vec{\w}^2+2\ba\s{\vec{k}}{\vec{\w}}
-\ba\vec{k}^2-\ba m^2\right\}}.
\ee
Applying similar reasoning as in Ref.~\cite{Watson:2007mz}, we come to the 
following parametric form of the integral (recall that 
$x=k_4^2$, $y=\vec{k}^2$):
\bea
A_m&=&\frac{(x+y+m^2)^{-1-\e}}{(4\pi)^{2-\e}}\G(1+\e)\int_0^1 d\ba
\int_0^{1-\ba}\,\frac{d\al}{ (\al+\ba)^{1/2}}
\left[\frac{\al\ba}{(\al+\ba)}
\frac{x}{(x+y+m^2)}+\frac{\ba(1-\ba)y+\ba m^2}{x+y+m^2}\right]^{-1-\e}.
\nonumber\\
\eea
For general values of $x$, the integral above cannot be solved because of 
the highly nontrivial denominator factor. Since there can be no 
singularities at $x=0$ (this would invalidate the Wick rotation which, as 
discussed in the text, does hold here), we make an expansion around this 
point and then integrate.  To first order in powers of $x$ we have:
\bea
A_m&\stackrel{x\rightarrow0}{=}&\frac{(x+y+m^2)^{-1-\e}}{(4\pi)^{2-\e}}
\G(1+\e)\int_0^1 d\ba\int_0^{1-\ba}d\al\,(\al+\ba)^{-1/2}
\left\{\left[\ba \frac{m^2+y(1-\ba)}{m^2+y}\right]^{-1-\e}\right.
\nonumber\\
&-&\left.\left[\ba \frac{m^2+y(1-\ba)}{m^2+y}\right]^{-2-\e}
\left[ \frac{\al\ba}{(m^2+y)(\al+\ba)}
-\ba \frac{m^2+y(1-\ba)}{(m^2+y)^2}\right](1+\e)x
\frac{}{}+{\cal O}(x^2)\right\}.
\eea
After performing the integration we get:
\bea
A_m(x,y)&\stackrel{x\rightarrow0}{=}&
\frac{(x+y+m^2)^{-1-\e}}{(4\pi)^{2-\e}}
(-2)\left\{-\frac{x}{m^2+y}
+\frac1\e \G(1+\e){}_{2}F_1\left(-\e,2+\e;1-\e;-\frac{y}{m^2+y}\right)
\right.\nonumber\\
&&\left.+\sqrt{1+\frac{m^2}{y}}
\ln\left(\frac{\sqrt{1+\frac{m^2}{y}}+1}{\sqrt{1+\frac{m^2}{y}}-1}\right)
+{\cal O}(x^2)
+{\cal O}(\e)
\right\}.
\label{eq:a4schw}
\eea
In the above formula, we isolated the hypergeometric term and evaluate  the 
$\e$ expansion  separately.  In order to do this, we have to differentiate 
the hypergeometric function with respect to the parameters.  In general, 
differentiation of ${}_{2}F_1(a,b;c;z)$ with respect to, e.g. the parameter 
$b$, gives (similar expressions are obtained for differentiation with 
respect to $a,c$):
\bea
{}_{2}F_1^{(0,1,0,0)}\left(a,b;c;z\right)&=&
\sum_{k=0}^{\infty}\frac{(a)_k (b)_k \Psi(b+k)}{(c)_k}\frac {z^k}{k!}
-\Psi(b){}_{2}F_1\left(a,b;c;z\right),
\label{eq:hypexp}
\eea
where $\Psi(k)$ is the digamma function and the Pochhammer symbol 
$(a)_k=\G(a+k)/\G(a)$ (see, for instance, \cite{bateman}).  With the help of 
formula \eq{eq:hypexp}, we get:
\bea
{}_{2}F_1\left(-\e,2+\e;1-\e;z\right)=1+\e\ln(1-z)+{\cal O}(\e).
\eea
Inserting this back into \eq{eq:a4schw}, we can write down the result for 
the integral $A_m$ (to first order in powers of $x$):
\bea
A_m(x,y)&\stackrel{x\rightarrow0}{=}&\frac{(x+y+m^2)^{-1-\e}}{(4\pi)^{2-\e}}
\left\{-\frac{2}{\e}+2\ga+2\left[-\ln\left(\frac{m^2}{m^2+y}\right)
+\frac{x}{m^2+y}
\right]\right.\nonumber\\
&&\left.-2\sqrt{1+\frac{m^2}{y}}
\ln\left(\frac{\sqrt{1+\frac{m^2}{y}}+1}{\sqrt{1+\frac{m^2}{y}}-1}\right)
+{\cal O}(x^2)+\cal O(\e)\right\},
\eea
which agrees explicitly with the corresponding expansion of the result given 
in \eq{eq:asol}.  

We now turn to the integral $A^4_m$, given by \eq{eq:a4def}. The parametric 
form has the expression:  
\bea
A^4_m&=&k_4\frac{(x+y+m^2)^{-1-\e}\G(1+\e)}{(4\pi)^{2-\e}}\int_0^1 d\ba
\int_0^{1-\ba}\!\!d\al\,\frac{\ba}{(\al+\ba)^{3/2}}
\left[\frac{\al\ba}{(\al+\ba)}
\frac{x}{(x+y+m^2)}
+\frac{\ba(1-\ba)y+\ba m^2}{x+y+m^2}\right]^{-1-\e}.
\eea
Calculations similar to the integral $A_m$ bring us to the following result 
(to first order in $x$):
\bea
A^4_m(x,y)&\stackrel{x\rightarrow0}{=}&
k_4\frac{(x+y+m^2)^{-1-\e}}{(4\pi)^{2-\e}}
\left\{
2\sqrt{1+\frac{m^2}{y}}
\ln\left(\frac{\sqrt{1+\frac{m^2}{y}}+1}{\sqrt{1+\frac{m^2}{y}}-1}\right)
+2\left(1+\frac{m^2}{y}\right)
\ln\left(\frac{m^2}{m^2+y}\right)\right.\nonumber\\
&&\left.-\frac23\frac{x}{y}\left[
1+\left(\frac{m^2}{y}-2\right)\ln\frac{m^2}{m^2+y}-
\frac{2}{\sqrt{1+\frac{m^2}{y}}}
\ln\left(\frac{\sqrt{1+\frac{m^2}{y}}+1}{\sqrt{1+\frac{m^2}{y}}-1}\right)
\right]
+{\cal O}(x^2)
+{\cal O}(\e)\right\},
\eea
again in agreement with the expansion of the result given in \eq{eq:a4sol}. 

Another useful check comes from the study of the mass differential 
equation.  With $I^n$ given by \eq{eq:idef}, the derivative with respect to 
the mass reads:
\be
m\frac{\pd I^n}{\pd m}=
\int\frac{\dk{\w}\,\w_4^n}{\w^2[(k-\w)^2+m^2]\vec{\w}^2}
\left\{-2\frac{m^2}{(k-\w)^2+m^2}\right\}.
\label{eq:ademass}
\ee
From the relations \eq{eq:ade0}, \eq{eq:adevec} and \eq{eq:ademass} we get 
the following relation:
\be
k_4\frac{\pd I^n}{\pd k_4}
+k_k\frac{\pd I^n}{\pd k_k}+m\frac{\pd I^n}{\pd m}=(d+n-5)I^n.
\ee
Using the same procedures as in the text, we can then derive a differential 
equation for the integral in terms of the mass:
\be
m^2\frac{\pd I^n}{\pd m^2}=(d+n-4)\frac{m^2}{k^2+m^2}I^n
-\frac{m^2}{k^2+m^2}\int\frac{\dk{\w}\,\w_4^n}{[(k-\w)^2+m^2]^2\vec{\w}^2}.
\label{eq:adeint}
\ee

Starting with the case $n=0$ where $I^0\equiv A_m$, we see that by 
inserting the solution, \eq{eq:asol}, we have that in the limit 
$\e\rightarrow0$
\be
m^2\frac{\pd A_m}{\pd m^2}+(1+2\e)\frac{m^2}{k^2+m^2}A_m
=-\frac{m^2(x+y+m^2)^{-2-\e}}{(4\pi)^{2-\e}}
\frac{1}{y\sqrt{1+\frac{m^2}{y}}}
\ln\left(\frac{\sqrt{1+\frac{m^2}{y}}+1}{\sqrt{1+\frac{m^2}{y}}-1}\right)
+{\cal O}(\e).
\label{eq:masszero1}
\ee
In terms of Schwinger parameters, the explicit integral of \eq{eq:adeint} 
reads:
\be
-\frac{m^2}{k^2+m^2}\int\frac{\dk{\w}}{[(k-\w)^2+m^2]^2\vec{\w}^2}
=-\frac{m^2}{x+y+m^2}\frac{\G(1+\e)}{(4\pi)^{2-\e}}
\int_0^1d\al\,(1-\al)^{-1/2-\e}(m^2+\al y)^{-1-\e}
\label{eq:masszero2}
\ee
and for $m^2\neq0$ indeed
\be
-\frac{m^2}{k^2+m^2}\int\frac{\dk{\w}}{[(k-\w)^2+m^2]^2\vec{\w}^2}
=-\frac{m^2(x+y+m^2)^{-2-\e}}{(4\pi)^{2-\e}}
\frac{1}{y\sqrt{1+\frac{m^2}{y}}}
\ln\left(\frac{\sqrt{1+\frac{m^2}{y}}+1}{\sqrt{1+\frac{m^2}{y}}-1}\right)
+{\cal O}(\e),
\ee
showing that the mass differential equation is satisfied.  For $m^2=0$, the 
right-hand side of \eq{eq:masszero1} vanishes as $m^2\ln m^2$, whereas the 
parametric form of the integral in \eq{eq:masszero2} goes like $m^2/\e$.  
The integral of \eq{eq:masszero2} does contain an ambiguity in the ordering 
of the limits $m^2\rightarrow0$ and $\e\rightarrow0$, but this problem is 
not of importance because of the overall factor $m^2$ in the differential 
equation.  In fact, since the solution of the mass differential equation is 
in principle formally derived as the integral over $m^2$ and $m^2=0$ is the 
only the limit of this integral, the ambiguity encountered may be regarded 
as an integrable singularity and presents no problem.

Turning now to the case $n=1$ where $I^1\equiv A_m^4$, we first extract the 
overall $k_4$ factor as before by defining $A^4=k_4\ov{A}_m$ such that the 
differential equation is
\be
m^2\frac{\pd\ov{A}_m}{\pd m^2}=-2\e\frac{m^2}{k^2+m^2}\ov{A}_m
-\frac{m^2}{k^2+m^2}\int\frac{\dk{\w}}{[(k-\w)^2+m^2]^2\vec{\w}^2}.
\label{eq:masszero3}
\ee
Notice that in the integral term we have used the identities
\be
\int\frac{\dk{\w}\,\w_4}{[(k-\w)^2+m^2]^2\vec{\w}^2}=
\int\frac{\dk{\w}\,(k_4-\w_4)}{[\w^2+m^2]^2\left(\vec{k}-\vec{\w}\right)^2}
=k_4\int\frac{\dk{\w}}{[(k-\w)^2+m^2]^2\vec{\w}^2}.
\ee
Now, for $m^2\neq0$, the integral term of \eq{eq:masszero3} is finite as 
$\e\rightarrow0$; however, the $m^2=0$ limit is again ambiguous but as 
above this can be regarded as an integrable singularity.  Also, when 
$m^2=0$, $\ov{A}$ is known to be $\e$ finite (it is the massless integral 
considered in Ref.~\cite{Watson:2007mz}).  This means that as 
$\e\rightarrow0$ we have the simple integral expression
\be
m^2\frac{\pd\ov{A}_m}{\pd m^2}=-\frac{1}{x+y+m^2}\frac{1}{(4\pi)^{2-\e}}
\frac{1}{y\sqrt{1+\frac{m^2}{y}}}
\ln\left(\frac{\sqrt{1+\frac{m^2}{y}}+1}{\sqrt{1+\frac{m^2}{y}}-1}\right).
\ee
Knowing the solution, \eq{eq:a4sol}, it suffices to show that when 
$m^2=0$ the original massless integral from Ref.~\cite{Watson:2007mz} is 
reproduced and that the derivative of the massive solution satisfies the 
above.  Both of these steps are straightforward.

%-----------------------------------------------------------------------------

%-----------------------------------------------------------------------------

%-----------------------------------------------------------------------------
\end{document}